\documentclass[12pt]{iopart}
\usepackage{iopams}
\usepackage{bbm}
\usepackage{graphicx}
\usepackage[a4paper,left=2.5cm,right=2.5cm,top=\dimexpr15mm+1.5\baselineskip,bottom=2cm]{geometry}
\usepackage{esvect}
\usepackage{color,cite}
\usepackage{bm}
\usepackage{amsfonts, amssymb, pgfplots, float, mathtools, physics }
\pgfplotsset{compat=newest,compat/show suggested version=false}
\usepackage{color}
\usepackage{geometry}
\usepackage{mathdots}
\usepackage{tikz}
\usepackage[colorlinks=true,linkcolor=blue,citecolor=blue, urlcolor=blue]{hyperref}
\usepackage{comment}
\usepackage[normalem]{ulem}
\DeclareUnicodeCharacter{2212}{-}
\usepackage{dcolumn}
\usepackage{ulem}
\usepackage{tabularx}
\usepackage{colortbl,xcolor}

\begin{document}
\title{Anomalous current fluctuations and mobility-driven clustering}
\author{Tanmoy Chakraborty$^{1,2}$ and Punyabrata Pradhan$^{1}$}

\address{$^1$Department of Physics of Complex Systems, S. N. Bose National Centre for Basic Sciences, Block $-$ JD, Sector $-$ III, Salt Lake, Kolkata, 700106, India.}
\address{$^2$Department of Physics, Technion $-$ Israel Institute of Technology, Haifa, 3200003, Israel.}

\ead{tanmoy.statphys@gmail.com}

\begin{abstract}
We study steady-state current fluctuations in hardcore lattice gases on a ring of $L$ sites, where $N$ particles perform symmetric, {\it extended-ranged} hopping. The hop length is a random variable depending on a length scale $l_0$ (hopping range) and the inter-particle gap. The systems have mass-conserving dynamics with global density $\rho = N/L$ fixed, but violate detailed balance. We consider two analytically tractable cases: (i) $l_0 = 2$ (finite-ranged) and (ii) $l_0 \to \infty$ (infinite-ranged); in the latter, the system undergoes a clustering or condensation transition below a critical density $\rho_c$. In the steady state, we compute, exactly within a closure scheme, the variance $\langle Q^2(T) \rangle_c = \langle Q^2(T) \rangle - \langle Q(T) \rangle^2$ of the cumulative (time-integrated) current $Q(T)$ across a bond $(i,i+1)$ over a time interval $[0, T]$. We show that for $l_0 \to \infty$, the scaled variance of the time-integrated bond current—or, equivalently, the mobility—diverges at $\rho_c$. That is, near criticality, the mobility $\chi(\rho) = \lim_{L \to \infty} [\lim_{T \to \infty} L \langle Q^2(T, L) \rangle_c / 2T] \sim (\rho - \rho_c)^{-1}$ has a simple-pole singularity, thus providing a dynamical characterization of the condensation transition, previously observed in a related mass aggregation model by Majumdar et al.\ [{\it Phys.\ Rev.\ Lett.\ {\bf 81}, 3691 (1998)}]. At the critical point $\rho = \rho_c$, the variance has a scaling form $\langle Q^2(T, L) \rangle_c = L^{\gamma} {\cal W}(T/L^{z})$ with $\gamma = 4/3$ and the dynamical exponent $z = 2$. Thus, near criticality, the mobility {\it diverges} while the diffusion coefficient remains {\it finite}, {\it unlike} in equilibrium systems with short-ranged hopping, where diffusion coefficient usually {\it vanishes} and mobility remains finite.

\end{abstract}

\section{Introduction}
Characterizing the large-scale spatio-temporal properties of many-body systems \cite{Spohn_2012}, particularly those undergoing a phase transition in an out-of-equilibrium setting \cite{Marro_Dickman_1999}, remains a formidable challenge in statistical physics. Over the years, it has inspired physicists to extend conventional equilibrium hydrodynamic theories, e.g., those formulated by Hohenberg and Halperin \cite{Halperin_RMP}, to nonequilibrium regimes. However, one may wonder if such a theory can always be applied in a nonequilibrium scenario. If not, when does such approach break down? 
A particularly interesting case is that of the clustering or condensation transitions observed in mass-conserving systems. Such transitions refer to the phenomenon in which a system undergoes a \textcolor{black}{macroscopic transformation} upon tuning a parameter, such as global density. They are ubiquitous in nature and usually characterized through the bulk-(collective-)diffusion coefficient (or, equivalently, the relaxation rate), which vanishes at the critical point, leading to what is known as {\it critical slowing down} \cite{Halperin_RMP, Kawasaki_csd, Binder_CSD, Bhattacharjee_1985}. 
However, in nonequilibrium, that scenario may not always hold true. 
Indeed, in this paper, \textcolor{black}{we construct a model,} where there is no dynamical slowing down at the critical point, and instead the transition is characterized by an instability (divergence) in the current fluctuations. To this end, we consider a class of paradigmatic lattice gases with {\it extended-ranged} hopping, where the hopping range is characterized by a length scale $l_0$. \textcolor{black}{In the case of {\it infinite-ranged} hopping,} the system undergoes a nonequilibrium condensation transition below a critical density $\rho_c$. Quite remarkably, at the critical density, unlike in equilibrium, we show that, while the bulk-diffusion coefficient \textcolor{black}{remains finite, the mobility,} characterizing the current fluctuations, {\it diverges}.

To illustrate the above point, let us consider a one-dimensional system with mass-conserving, diffusive dynamics on a periodic domain of size $L$. The time evolution of the density field $\rho(x, \tau)$ at suitably scaled (hydrodynamic) position $x$ and time $\tau$ is described by the following fluctuating hydrodynamic equation \cite{MFT-RMP2015}:
\begin{equation}
\label{fluctuating_hydrodynamics}
\frac{\partial \rho(x, \tau)}{\partial \tau} = - \frac{\partial}{\partial x} \left[ - D(\rho) \frac{\partial \rho}{\partial x} + \sqrt{\frac{2 \chi(\rho)}{L} } \zeta(x,\tau)\right] \equiv - \partial_x {\cal J}(\rho).
\end{equation} 
Eq.~\eqref{fluctuating_hydrodynamics} is a continuity equation, where the coarse-grained current ${\cal J}(\rho) = J^{(D)}(\rho) + J^{(fl)}(\rho)$ in the square bracket on the right-hand side consists of two components. The first component $J^{(D)}$ of the current characterizes the relaxation of density perturbations towards the steady state with $D(\rho)$ denoting the collective-(bulk-)diffusion coefficient. The second one $J^{(fl)}(\rho)$ accounts for the fluctuating, or the ``noise'', current characterized by the density-dependent Onsager transport coefficient $\chi(\rho)$ and $\zeta(x, \tau)$, a Gaussian white noise, \textcolor{black}{uncorrelated in space and time and having zero mean and unit variance.} 
More specifically, in a one-dimensional diffusive system on a lattice of $L$ sites and with densities $\rho$ and $\rho+\Delta \rho$ at the left and right boundaries, respectively, the average steady-state current behaves as \cite{Derrida_PRL2004}
 \begin{equation}
     \lim_{t \to \infty} \frac{\langle Q_i(t) \rangle}{t} = D(\rho) \frac{\Delta \rho}{L}, \label{mean1}
 \end{equation}
 where  $Q_i(t)$ is the time-integrated current across a bond $(i,i+1)$ in (microscopic) time interval $t \gg L^2$ and $D(\rho)$ is the bulk-diffusion coefficient.
 The above phenomenological relation, characterizing the response of the system to a density gradient, is known as Fick's law. 
On the other hand, on a periodic domain ($\Delta \rho=0$) where the net current vanishes in the system, the steady-state variance or the second cumulant $\langle Q^2_i(t) \rangle_c = \langle Q^2_i(t) \rangle - \langle Q_i(t) \rangle^2$ of time-integrated bond current in the steady state behaves as  \cite{Derrida_PRL2004}
\begin{equation}
 \lim_{t \to \infty} \frac{\langle Q^2_i(t) \rangle_c}{t} = \frac{2\chi(\rho)}{L}, \label{var1}
   \end{equation}
where the density-dependent quantity $\chi(\rho)$ is called the Onsager transport coefficient or the mobility. 
However, explicit calculations of the two transport coefficients remain a challenge, especially for interacting systems, which have nontrivial many-body correlations and can exhibit a phase transition.
Previously, in Ref. \cite{tanmoy_2020}, we calculated the bulk-diffusion coefficient for such interacting-particle systems of hardcore lattice gases having extended-range hopping.
In this paper, we analytically calculate the mobility by directly calculating the variance of time-integrated current as in Eq. \eqref{var1} in these systems. Thus, we characterize the condensation transition (observed when hopping range becomes infinite) by showing that the mobility in that case has a simple-pole singularity $\chi(\rho) \sim (\rho - \rho_c)^{-1}$ near criticality ($\rho \to \rho_c^+$).  Notably, unlike in equilibrium, the bulk-diffusion coefficient $D(\rho)$ at the critical point \textcolor{black}{does not vanish.}

The time-dependent properties of interacting-particle systems have been of significant interest in the past decades \cite{Halperin_RMP}. In many cases where there is a conserved quantity (local density), the quantities of interest have been current fluctuations and the higher-order cumulants. Indeed, current fluctuations have been studied in a variety of models, including simple exclusion processes \cite{Derrida:1998, Derrida_PRL2004, Derrida:2007, Derrida-PRE2008, Derrida_2009_JSTAT_1, Imparato:2009, Krapivsky:2012}, contact processes \cite{Gorissen:2009}, zero-range processes \cite{Harris:2005}, models of heat conduction \cite{Kipnis:1982, Basile:2006, Hurtado:2009}, and, more recently, active matter systems \cite{GrandPre:2018, Banerjee:2020, Chakraborty_PRE_2024_RTP_II, Jose_PRE_2023}, using microscopic theory, Monte Carlo simulations, or in the hydrodynamic framework of macroscopic fluctuation theory \cite{Bertini_PRL2001, Bertini:2002, MFT-RMP2015, Derrida:2007}. 
However, despite their interesting behaviors, they have usually been studied far from a phase transition point, if any. 
As stated before, our main focus here is to study the \textcolor{black}{dynamic fluctuations near criticality in} systems exhibiting a nonequilibrium condensation transition.
Indeed, over the past decades, considerable progress has been made in understanding dynamic fluctuations in systems undergoing a phase transition, where several authors explore the anomalous temporal growth of the current fluctuation and its finite-size scaling at the transition point. For example, in the case of the ABC model exhibiting a clustering transition \cite{Evans_PRL_1998}, the scaled fluctuation of time-integrated current at criticality \textit{diverges} algebraically with system size \cite{Gerschenfeld_EPL_2011}. System-size-dependent anomalous (algebraic) growth of current fluctuations has also been characterized in certain chemical reaction networks \cite{Remlein_JCP_2024, Fiore:2021}, biochemical oscillators \cite{Nguyen_JCP_2018}, and more recently, in the nonequilibrium Curie-Weiss model \cite{ptaszynski_PRE_2024}, where the heat-current fluctuations are shown to diverge algebraically near a temperature-driven transition point and exponentially near a field-induced transition. Apart from the anomalous system-size dependence, recent studies have also revealed an unusual temporal growth of current fluctuations near a phase transition point, e.g., absorbing-phase transition in conserved (``fixed-energy'') sandpiles \cite{Anirban-PRE2023, Mukherjee_PRE_2024} and condensation transition in equilibrium systems such as symmetric zero-range processes \cite{Chakraborty_PRE_2024_ZRP}.
Interestingly, nonanalyticity in current fluctuations at the critical point has been reported in a certain class of active lattice gases exhibiting the motility-induced phase separation (MIPS) \cite{agranov_2023_scipost, Jose:2023}.

In this paper, we investigate current fluctuations in extended-range lattice gases. In particular, we explore the precise dynamical mechanism behind the nonequilibrium condensation transition, which is observed by tuning the hopping range to infinity (i.e., infinite-ranged hopping) and global density below critical density $\rho_c$. To this end, we theoretically study the steady-state variance or the second cumulant of time-integrated currents across a single bond as well as the entire system, both near and far from criticality, and we characterize the current fluctuations in terms of the Onsager coefficient or the mobility as defined in Eq. \eqref{var1}. 
\textcolor{black}{We introduce a closure approximation scheme,} which breaks the infinite BBGKY hierarchy involving the $n-$point correlations ($n \ge 3$) and enables us to explicitly calculate the two-point correlations for current and density and thus to obtain the closed-form analytic expressions for time-integrated current fluctuations in the two limiting cases of finite-ranged ($l_0=2$) and infinite-ranged ($l_0 \to \infty$) hopping. 
In the former case and in the latter case, when the system is far from criticality, where the transport coefficients remain finite, the steady-state variance $\langle Q^2_i(t) \rangle_c = \langle Q^2_i(t) \rangle$ or the second moment (as $\langle Q_i(t) \rangle=0$ in steady state) of time-integrated bond current follows the behavior typically observed in a diffusive system: it crosses over from a subdiffusive growth $\sqrt{t}$ at intermediate, \textcolor{black}{ but still large,} times ($1 \ll t \ll L^2$) to a diffusive one $t/L$ at long times ($t \gg L^2$) \cite{Derrida-Sadhu-JSTAT2016}, with both the regimes found to be remarkably connected through a single scaling function \cite{Chakraborty_PRE_2024_RTP_II,  Hazra_Jstat_2024}. Interestingly, for infinite-range hopping, at the onset of the condensation transition, the system shows an anomalously enhanced and qualitatively different growth regime for the time-integrated bond-current fluctuations, thus marking a departure from the conventional growth behavior.
Notably, the enhancement of current fluctuation across bonds leads to a power-law divergence $\sim (\rho-\rho_c)^{-1}$ in the scaled fluctuation of space-time integrated currents across the entire system - the phenomenon we call {\it mobility-driven clustering}. 
We show that, at the critical point, where the mobility diverges, the current fluctuations greatly increase as compared to those observed far from criticality. Indeed, using a scaling theory, we show that it crosses over from an anomalously suppressed (subdiffusive), yet larger-than-typical \cite{Derrida-Sadhu-JSTAT2016,  Hazra_Jstat_2024}, $t^{2/3}$ growth regime to a linear $t/L^{2/3}$ one, \textcolor{black}{albeit with an anomalous system-size-dependent prefactor. }
We substantiate our theoretical predictions through Monte Carlo simulations of the models.

\section{Model}
\label{sec-ch3-model}

We now define the model, which consists of $N$ hardcore particles diffusing on a one-dimensional periodic lattice of $L$ sites. The hardcore constraint sets the maximum number of particles occupying a lattice site to one and prohibits particle crossing (i.e., single-file motion). The system is governed by a continuous-time Markov process, where a particle hops out with unit rate in either direction with equal probability $1/2$, where the hop length ${\rm min} \{g, l\}$ with $g$ being the gap size along the hopping direction and $l$ is drawn from a distribution $\phi(l)$ having a characteristic length scale $l_0$. 
 That is, the particle hops by length $l$ if the empty lane (of consecutive vacancies or holes) having the inter-particle {\it gap} of size $g$ (lane size), along the hopping direction, exceeds $l$; otherwise, due to the hardcore constraint, the particle moves through the entire gap $g$ and resides just beside the next particle along its hopping direction.

For analytical tractability, we simply choose the following distribution $\phi(l) = \alpha \delta_{l,1} + \beta \delta_{l,l_0}$ consisting of short-ranged and long-ranged particle hopping with $l_0>1$ and probabilities $\alpha$ and $\beta$, respectively, with $\alpha + \beta = 1$. Therefore, during a hopping event, one of the following events occurs with the respective probability. 
\\
\\
(A) {\it Short-ranged hop.$-$} A particle attempts a short-ranged hop, with probability $\alpha$. During the particle-transfer, it goes to its right or left nearest neighboring site with equal probability $1/2$, given that the destination site is {\it unoccupied}. \\
\\
(B) {\it Extended-ranged hop.$-$} A particle attempts an extended-ranged hop with probability $\beta$ to the left or right with equal probability $1/2$. If $g \ge l_0$, the particle hops by length $l_0$; otherwise, it travels the entire gap in the hopping direction and sits beside the nearest particle. 
\\\\
Depending on the characteristic hop length $l_0$ being finite or infinite, we can categorize the models into two classes: finite-ranged hopping (FRH) and infinite-ranged hopping (IRH), respectively. Essentially, the latter implies that, during an extended-ranged hop, a particle travels the entire gap in the hopping direction and then sits beside the nearest occupied site. 
In a special case, when only the short-ranged hopping (A) is present (i.e., $\alpha=1$), the microscopic dynamics would simply boil down to that of symmetric simple exclusion processes (SSEP) \cite{Derrida_2009_JSTAT_1}. 
Also, in the case of IRH, we have $0 < \alpha < 1$ and then the extended-ranged hopping (B) happens with a nonzero probability  $\beta=1-\alpha$. Physically, it is evident that the infinite-ranged hopping promotes vacancy clustering, since it could effectively merge vacancies on both sides of the hopping particle. On the other hand, the short-ranged hopping tends to hinder the clustering process. Consequently, the two dynamical update rules compete with each other and, below a critical density, infinite-ranged hopping becomes dominant, leading to a condensation transition characterized by vacancy clustering. 
\textcolor{black}{Notably, the infinite-ranged model is related to a conserved-mass aggregation model on a periodic ring. In the latter system, particle or mass occupancy at a lattice site is unbounded and the dynamics involve aggregation and fragmentation (chipping of a single unit) of masses via symmetric nearest neighbor hopping \cite{Barma_PRL_1998}, and the existence of clustering transition was reported  there. Indeed, the infinite-ranged hopping studied in this paper mimics the aggregation dynamics in the unbounded mass aggregation model, where an entire ``chunk" of mass hops out from a site and coalesces with the mass at any one of the nearest-neighbor sites. On the contrary, the particle transfer corresponding to the short-ranged hop is analogous to the fragmentation (or chipping), where a single unit of mass breaks off from a site and then gets deposited at one of the nearest neighbors with equal probability.}

\section{Theory for bond-current fluctuation }
\label{sec:ch4-theory}

In this section, we use a truncation (approximate) scheme and develop a microscopic dynamical theory to analytically calculate current fluctuations in the LLGs as defined in Sec.~\ref{sec-ch3-model}. We then substantiate the analytic results with direct Monte-Carlo simulations of the models.

\subsection{Average bond-current}\label{sec:ch4-av_current}

To begin with, let us first define the time-integrated or cumulative current $Q_{i}(t)$ across a bond $[i, i+1]$, which quantifies the net particle flux across the bond up to time $t$. 
Notably, in the case of extended-ranged hopping, a particle contributes to the current across all bonds along its hopping direction up to the destination site. In other words, when a particle hops from site $i$ to $j$ along right (left) directions, it contributes a unit increment (decrement) of cumulative current across all the bonds in between.
The cumulative bond current $Q_i(t)$ is a time-extensive quantity that can be easily measured in simulations. On the other hand, the corresponding instantaneous bond current $J_i(t)$, which is a series of delta peaks, can be defined as follows:
\begin{eqnarray}\label{ch4-inst_and_integrated_current}
J_{i}(t) \equiv  \lim_{\Delta t \rightarrow 0} \frac{\Delta Q_i }{\Delta t},
\end{eqnarray}
where $\Delta Q_i(t)=\int_{t}^{t+\Delta t } J_{i}(t) dt $ is the time-integrated bond-current in the time interval $\Delta t$. 
Note that, as the model dynamics incorporates particle hopping that is symmetric or unbiased, we expect that $\langle J_{i}(t) \rangle$ will be generated solely by the density gradient, leading us to determine the bulk-diffusion coefficient $D(\rho)$ in the system from the Fick's law. Before delving into the calculation details, we now define the following stochastic variables,
  \begin{eqnarray}
 \mathcal{U}_{i+l}^{(l)} &\equiv& \overline{\eta}_{i+1} \overline{\eta}_{i+2} \dots \overline{\eta}_{i+l} , \\ 
 \mathcal{V}_{i+l+1}^{(l+2)} &\equiv& \eta_{i}\overline{\eta}_{i+1} \overline{\eta}_{i+2} \dots \overline{\eta}_{i+l}\eta_{i+l+1} , 
 \end{eqnarray}
 where $\bar{\eta}_i=(1-\eta_i)$,  $\mathcal{U}^{(l)}$ and $\mathcal{V}^{(l+2)}$ are indicator functions for a single site being vacant, $l$ consecutive sites being vacant and a vacancy cluster to be of size $l$, respectively. 
 Notably, for a particle moving rightward (leftward) across a bond, current across the bond is increased (decreased) by unity.
 In the subsequent discussions, we follow the formalism developed previously in Ref. \cite{Chakraborty_PRE_2024_RTP_II}. 
 However, unlike the systems studied in Ref. \cite{Chakraborty_PRE_2024_RTP_II}, the systems considered in this paper can undergo a phase transition and, quite remarkably, admit a closed-form analytic solution for the current fluctuations (as an explicit function of time, density, system sizes, and other parameters), which is in general a quite challenging task for an interacting-particle system.

 The continuous-time evolution for time-integrated current $Q_i(t)$ in an infinitesimal time interval $[t, t+dt]$ can be written as
\begin{eqnarray} 
 Q_i(t+dt) = 
\left\{
\begin{array}{ll}
\vspace{0.15 cm}
 Q_i(t)  + 1,            ~~~  & {\rm prob.}~~~ \mathcal{P}^{R}_{i}(t) dt , \\
\vspace{0.15 cm}
 Q_i(t) - 1,            ~~~  & {\rm prob.}~~~ \mathcal{P}^{L}_{i}(t) dt, \\
 \vspace{0.15 cm}
 Q_i(t),                ~~~  & {\rm prob.}~~~~  1 - (\mathcal{P}^{R}_{i} + \mathcal{P}^{L}_{i}) dt, \\
\end{array}
\right.
\label{ch4-Q_update_eq}
\end{eqnarray}
where $\mathcal{P}^{R}_{i} dt$ and $\mathcal{P}^{L}_{i} dt$ are probabilities of the hopping events which we will determine now. To compute the hopping rate $\mathcal{P}^{R}_{i}$, we consider below all possible rightward hopping events which generate unit increment in $Q_i$  across the bond $[i, i+1]$.  
\\
\paragraph*{Short-ranged hop:} With rate $\alpha$, a particle from the site $i$ hops symmetrically to the right neighboring site, i.e., $i+1$, provided the site is unoccupied. In this case, the corresponding probability term is given by $\mathcal{P}^{R, sh}_{i} dt$, where
\begin{eqnarray}\label{ch4-hopping_rate_sh}
\mathcal{P}^{R, sh}_{i} = \frac{\alpha}{2} \eta_{i} (1-\eta_{i+1}) = \frac{\alpha}{2} \left(\eta_i - \mathcal{V}^{(2)}_{i+1}\right).
\end{eqnarray}
\paragraph*{Extended-ranged hop:} In this case, particle symmetrically hops with rate $\beta$ and depending on the gap size $g$ along the hopping direction and the attempted hop-length $l_0$, we consider the following two cases:

\begin{itemize}
\item[I.] $g < l_0:$ In this particular case of gap size being smaller than the attempted hopping length, a particle traverses the entire gap of length $g$ and crosses the bond $[i, i+1]$ only when the bond resides within the vacancy or gap cluster. We also realize that,
for a fixed $g$, current across a particular bond $[i, i+1]$ can be contributed by translating the entire cluster in $g$ possible ways. The corresponding contribution to
the probability is given by $P_{i}^{R,g < l_0} dt$ where,
\begin{eqnarray}\label{ch4-hopping_rate_lh_g<l}
P_{i}^{R,g < l_0} = \frac{\beta}{2}\sum_{k=1}^{g}\mathcal{V}^{(g+2)}_{i+k+1}.
\end{eqnarray}

\item[II.] $g \geq l_0:$ In the opposite scenario, the particle hops by length $l_0$, and for the contribution of $Q_i$ caused by the above move, the particle must cross the bond $[i, i+1]$. Following the preceding argument, we obtain current contribution across the bond $[i, i+1]$ in $l_0$ in different ways. The corresponding contribution to the probability is given by $P_{i}^{R,g \geq l_0} dt$ where,
\begin{eqnarray}\label{ch4-hopping_rate_lh_g>l}
P_{i}^{R,g \geq l_0} = \frac{\beta}{2}\sum_{k=1}^{l_0}\left(\mathcal{U}_{i+k}^{(l_0)} - \mathcal{U}_{i+k}^{(l_0+1)}\right).
\end{eqnarray}
\end{itemize}
Now, considering all possible gap sizes and using the above contributions in Eqs.~\eqref{ch4-hopping_rate_sh}, \eqref{ch4-hopping_rate_lh_g<l}, and, \eqref{ch4-hopping_rate_lh_g>l}, the total rightward hopping rate is calculated to be
\begin{eqnarray}
\label{ch4-pr}
  \mathcal{P}^{R}_{i} &\equiv& \frac{\beta}{2}\left[\sum_{k=1}^{l_0} \left(\mathcal{U}_{i+k}^{(l_0)} - \mathcal{U}_{i+k}^{(l_0+1)}\right) + \sum_{g=1}^{l_0-1}\sum_{k=1}^{g}\mathcal{V}_{i+k+1}^{(g+2)}\right]  + \frac{\alpha}{2}\left(\eta_{i}- \mathcal{V}^{(2)}_{i+1} \right).
\end{eqnarray}
Similarly, considering the leftward hopping events, we directly obtain the leftward hopping rate 
\begin{eqnarray}
\mathcal{P}^{L}_{i} &\equiv & \frac{\beta}{2} \left[\sum_{k=1}^{l_0} \left(\mathcal{U}_{i+k-1}^{(l_0)} - \mathcal{U}_{i+k}^{(l_0+1)}\right) + \sum_{g=1}^{l_0-1}\sum_{k=1}^{g}\mathcal{V}_{i+k}^{(g+2)}\right] + \frac{\alpha}{2}\left(\eta_{i+1}- \mathcal{V}^{(2)}_{i+1} \right).  \label{ch4-pl}
\end{eqnarray}
By using the above microscopic update rules and doing some straightforward algebraic manipulations, we have, by definition, the average instantaneous current 
\begin{equation}
    \left \langle J_{i}(t) \right \rangle \equiv \langle J_{i}^D(t)\rangle,
\end{equation}
where the local (stochastic) diffusive current is given by the following expression:
\begin{eqnarray}
  \label{ch4-time-derivative-int-current_2}
J^{D}_i &=&  \frac{\beta}{2} \Bigg[\sum_{g=1}^{l_0-1}\left( \mathcal{V}_{i+g+1}^{(g+2)} - \mathcal{V}_{i+1}^{(g+2)} \right) + \left( \mathcal{U}_{i+l_0}^{(l_0)} -\mathcal{U}_i^{(l_0)} \right)\Bigg] + \frac{\alpha}{2}(\eta_{i} - \eta_{i+1}) .
\end{eqnarray}
It is important to note that $\left \langle J_{i}(t) \right \rangle$ in Eq.~\eqref{ch4-time-derivative-int-current_2} is expressed as the gradient of the local observables $\langle \eta \rangle$, $\langle \mathcal{V}^{(g+2)} \rangle$, and, $\langle \mathcal{U}^{(l_0)}\rangle$, which clearly suggests that the system possess a  ``\textit{gradient} property'' \cite{Spohn_2012, Landim_1998, Mallick-PRE2014}. Now, at large time, we assume the system attains a \textit{local steady state} (analogous to a local-equilibrium hypothesis), which implies that the time evolution of the above (``fast'') observables is effectively governed by the evolution of the conserved local density field (a ``slow'' variable). In that case, using the Taylor's series expansion around the local density $\rho$, we can write the average instantaneous current in the form of a diffusive current in the following manner:
\begin{eqnarray}
  \label{ch4-cont-eqn}
\left \langle J_i^{D}(t) \right \rangle \simeq - D(\rho) [\langle \eta_{i+1}(t) \rangle - \langle \eta_i(t) \rangle]  
\end{eqnarray}
where the bulk-diffusion coefficient $D(\rho)$ is a function of the global density $\rho$ and is given by the following expression
\begin{eqnarray}
 D(\rho)&=& \frac{\alpha}{2}-\frac{\beta}{2}\frac{\partial}{\partial \rho} \left[\sum_{g=1}^{l_0-1} g \langle \mathcal{V}^{(g+2)} \rangle(\rho) + l_0 \langle \mathcal{U}^{(l_0)} \rangle (\rho) \right],\\
 &=& \frac{\alpha}{2} -\frac{\beta}{2}\frac{\partial}{\partial \rho}\left[\rho \left(\sum_{g=1}^{l_0-1} g P(g) + l_0\sum_{g=l_0}^{\infty} (g-l_0+1) P(g)\right) \right].
 \label{ch4-bulk-diffusivity-LH}
 \end{eqnarray}
Note that we have arrived at Eq.~\eqref{ch4-bulk-diffusivity-LH}, which is exactly expressed in terms of inter-particle gap distribution, by using the following identities,
\begin{eqnarray}\label{ch4_multicorrelator_pg}
\langle \mathcal{V}^{(g+2)}\rangle(\rho) &=&\rho P(g), \\
\langle \mathcal{U}^{(l)}\rangle(\rho) &=& \rho \sum_{g=l}^{\infty} (g-l+1) P(g),
\end{eqnarray} 
directly connecting the correlation functions $\langle \mathcal{V}^{(g+2)}\rangle(\rho)$ and $\langle \mathcal{U}^{(l)}\rangle(\rho)$ to the gap-distribution function $P(g)$; of course, the gap distribution $p(g)$ is density-dependent and can be explicitly calculated in the two special cases of hopping range $l_0=2$ and $l_0 \to \infty$. 
\textcolor{black}{We mention here that the above derivation of the bulk-diffusion coefficient is somewhat different from that given in Ref. \cite{tanmoy_2020}, where we had calculated it by deriving the time-evolution equation for density. In this paper, first we directly construct the current operator, as given in Eq.~\eqref{ch4-time-derivative-int-current_2}, and then calculate the average current, which immediately gives us the expression of the bulk-diffusion coefficient. Now we proceed to calculate various other quantities involving the current fluctuations, which are the main focus of this paper. }

\subsection{Dynamic correlations for bond current}
\label{Sec:ch5:truncation}

To begin with, we define the spatiotemporal correlation function for two stochastic variables $A_{r}(t')$ and $B_{0}(t)$ as follows:
\begin{eqnarray}\label{ch4-correlation_defn}
\mathcal{C}^{AB}_{r}(t',t) &=& \left\langle A_{r}(t') B_{0}(t) \right\rangle_{c}, \nonumber \\ &=& \left\langle A_{r}(t') B_{0}(t) \right\rangle - \left\langle A_{r}(t')  \right\rangle \left\langle  B_{0}(t) \right\rangle.
\end{eqnarray}
Let us now consider the two quantities, $Q_{r}(t')$ and $Q_{0}(t)$, which are stochastic time-integrated currents, measured across bonds $(r, r+1)$ and $(0, 1)$ up to times $t'$ and $t$ $(t' > t)$, respectively. In this section, we are now going to calculate the spatio-temporal correlation between them, i.e., $\mathcal{C}^{QQ}_{r}(t',t)$. For unequal times $t' > t$, we find that $\mathcal{C}^{QQ}_{r}(t',t)$ satisfies the following exact time-evolution equation \cite{Anirban-PRE2023},
\begin{eqnarray}
\frac{d}{dt'}\mathcal{C}^{QQ}_{r}(t',t)  = \left \langle J^{D}_{r}(t') Q_{0}(t)\right \rangle_{c},
\label{ch4-time-evolution-Q-Q-general1}
\end{eqnarray}
where $J^{D}_{r}$ as given in Eq.~\eqref{ch4-time-derivative-int-current_2} is the stochastic diffusive-current across $r$th bond at time $t$. Furthermore, using the microscopic update rules in Eq.~\eqref{ch4-Q_update_eq} for the time evolution of time-integrated bond currents, it can be shown that the equal-time ($t=t'$) correlation function $\mathcal{C}^{QQ}_{r}(t,t)$ satisfies the following equation:
\begin{eqnarray}
    \frac{d}{dt}\mathcal{C}^{QQ}_{r}(t,t)  = \Gamma_{r}(t) + \left \langle J^{D}_{r}(t) Q_{0}(t)\right \rangle_{c} + \left \langle J^{D}_{0}(t) Q_{r}(t)\right \rangle_{c},
\label{ch4-time-evolution-Q-Q-general1_eqtime}
\end{eqnarray}
where the quantity $\Gamma_{r}$ is given by
	\begin{eqnarray}
	\label{ch4-gamma_r}
	\hspace{-1.5 cm}
	\Gamma_{r} \hspace{0 cm}= \hspace{-0.05 cm} \rho\left[\alpha \Big\{1-P(0, t)\Big\} \delta_{r,0} + \hspace{-0.0 cm} \beta\sum_{l=\mid r \mid +1}^{\infty} \hspace{-0.25 cm} \delta_{l,l_0} \hspace{-0.0 cm} \Big\{(l-\mid r \mid) \sum_{g=l}^{\infty}P(g, t)  \hspace{-0 cm}+ \sum_{g=\mid r \mid}^{l-1} (g-\mid r \mid) P(g, t) \Big\} \right].~~~~
	\end{eqnarray}
    Note that both Eqs. \eqref{ch4-time-evolution-Q-Q-general1_eqtime} and \eqref{ch4-gamma_r} are exact; indeed, as we show later, the quantity $\Gamma_r$ is directly related to the density-dependent Onsager coefficient or the particle mobility $\chi(\rho)$, which enters into the fluctuating hydrodynamic time-evolution equation \eqref{fluctuating_hydrodynamics}.
 However, note that, to calculate the steady-state current fluctuations, we need to explicitly calculate $\Gamma_{r}$ through the steady-state gap distribution $P(g)$; this task will be performed in the two special cases later in section \ref{sec:ch4-results}.

Now, according to the expression of $J^{D}_{r}$ in Eq.~\eqref{ch4-time-derivative-int-current_2}, one finds that the calculation of two-point current-current correlation functions $\mathcal{C}^{QQ}_{r}(t',t)$ and $\mathcal{C}^{QQ}_{r}(t,t)$ requires calculations of two-point density-current correlations $\mathcal{C}^{\eta Q}_{r}(t',t)$ and two multi-point correlations $\mathcal{C}^{\mathcal{U}^{(l)}Q}_r(t',t)$ and $\mathcal{C}^{\mathcal{V}^{(g+2)}Q}_r(t',t)$, all of which are among the unknowns in the problem. However, one can immediately see that the time-evolution of $\mathcal{C}^{\mathcal{U}^{(l)}Q}_r(t',t)$ and $\mathcal{C}^{\mathcal{V}^{(g+2)}Q}_r(t',t)$ depends on the higher-order correlations, whose time evolution would then throw even higher-order terms, and so on. As a result, calculating these correlation functions involves solving an infinite hierarchy of equations that cannot be closed, thus making the exact determination of $\mathcal{C}^{QQ}_{r}(t',t)$ and $\mathcal{C}^{QQ}_{r}(t,t)$ not possible at this stage.
To address the aforementioned difficulty, here we resort to a truncation (closure) scheme that allows us to efficiently handle Eq.~\eqref{ch4-time-evolution-Q-Q-general1} and to calculate the desired quantities quite accurately. First, we note that, on the hydrodynamic (i.e., large space-time) scale where the fluctuations of density around the global steady-state profile are small, the non-conserved (``fast'') variables $\mathcal{V}^{(g+2)}$ and $\mathcal{U}^{(l)}$, appearing in the local stochastic diffusive-current $J_{r}^{D}$ in Eq.~\eqref{ch4-time-derivative-int-current_2}, can be assumed to be slave to local density (a ``slow'' variable), under a ``local-equilibrium-like'' scenario where the hydrodynamics should be valid. Consequently, their gradients can be written in terms of the gradient of the local density itself or, in this case, the gradient of the occupation variable as follows: 
\begin{eqnarray}
  \label{ch4-closure_approximation}
J^{D}_{r}(t') \simeq D(\rho) [ \eta_{r}(t') - \eta_{r+1}(t') ],
\end{eqnarray}
where $D(\rho)$ is the bulk-diffusion coefficient as given in Eq. \eqref{ch4-bulk-diffusivity-LH}.
The above equation could be thought of as Fick's law on a microscopic level; also, it is quite instructive to compare the above equation with Eq.~\eqref{ch4-cont-eqn}. Later, we explicitly calculate $D$ as a function of $\rho$ in the two cases $l_0=2$ and $\infty$ considered in this paper.
A straightforward consequence of Eq.~\eqref{ch4-closure_approximation} is that we can now simply replace the correlation between the stochastic diffusive current $J^{D}_{r}(t')$ and any arbitrary variable $B_{0}(t)$ as the gradient of correlations between the stochastic density and the variable $B_{0}(t)$. that is, we write the following correlation simply in terms of a two-point correlation,
\begin{eqnarray}
\left\langle J^{D}_{r}(t') B_{0}(t) \right \rangle_{c} \hspace{-0.0 cm} \simeq -D(\rho) \Delta_{r}\left\langle \eta_{r}(t') B_0(t)\right \rangle_{c},
\label{ch4-TS}
\end{eqnarray}
where $\Delta_{r} \eta_{r} = \eta_{r+1}-\eta_{r}$ is the discrete gradient of local density or occupation variable $\eta_{r}$.

\textit{Decomposition of current.} The ``gradient'' structure of local current \cite{Spohn_2012, MFT-RMP2015}, as evident in Eqs.~\eqref{ch4-time-derivative-int-current_2} and \eqref{ch4-cont-eqn}, implies that the average current $\left \langle J_{i}(t) \right \rangle$ is zero in the steady state (i.e., when the system achieves a spatially homogeneous density profile on a periodic domain). However, due to the inherent stochasticity involved in microscopic dynamical updates, we still expect a nonzero contribution in $J_{i}(t)$ on the level of fluctuations. Therefore, to appropriately incorporate fluctuations into the large-scale hydrodynamic theory, we resort to decomposing the current into  ``slow'' (hydrodynamic) and ``fast'' (``noise'') components,
\begin{eqnarray}
\label{ch4-current_decompose}
J_{i}(t) = J^{D}_{i}(t) + J^{fl}_{i}(t),
\end{eqnarray}
where $J^{D}_{i}(t)$ is a (still stochastic) local (average) diffusive current characterizing the slow (hydrodynamic) relaxation of local density and $J^{fl}_{i}(t)$ is the fluctuating, or the ``noise'', current. 
 
Physically, the fluctuating component $J^{fl}_{i}(t)$, which originates from the stochasticity in the microscopic dynamics, characterizes ``fast'' relaxations. Therefore, the identification of $J^{D}_i(t)$ as in Eq.~\eqref{ch4-time-derivative-int-current_2} immediately implies that the average fluctuating component is identically zero, i.e.,
\begin{equation}
\langle J^{fl}_{i}(t) \rangle =0.
\end{equation}
In fact, as derived later [see  Eq.~\eqref{ch4-fluc_current_correlation}], fluctuating (noise) current  $J^{fl}_{i}(t)$ is  indeed $delta$ correlated in time. 
However, it exhibits interesting spatial correlations and plays a crucial role in determining the dynamic characteristics of correlations for actual particle currents, which we do in the next section.

Now, following the truncation scheme in Eq.~\eqref{ch4-time-evolution-Q-Q-general1}, we immediately get rid of the unclosed multi-point correlators $\mathcal{C}^{\mathcal{U}^{(l)}Q}_r(t',t)$ and $\mathcal{C}^{\mathcal{V}^{(g+2)}Q}_r(t',t)$ and, the resulting time-evolution of $\mathcal{C}^{QQ}_{r}(t',t)$ reduces to the gradient of density-current correlation function $\mathcal{C}^{\eta Q}_{r}(t',t)$ as follows:
\begin{eqnarray}
\label{ch4-time-evolution-Q-Q-general_2}
\frac{d}{dt'}\mathcal{C}^{QQ}_{r}(t',t)  \simeq -D(\rho) \Delta_{r} \mathcal{C}^{\eta Q}_{r}(t',t).
\end{eqnarray}
Notably, the time evolution of $\mathcal{C}^{\eta Q}_{r}(t',t)$ can be easily shown to possess a closed structure, which has the following form:
\begin{eqnarray}\label{ch4-time-evolution-eta-Q-diff_time}
\frac{d}{dt'}\mathcal{C}^{\eta Q}_{r}(t',t)  = D(\rho) \Delta^{2}_{r} \mathcal{C}^{\eta Q}_{r}(t',t).
\end{eqnarray}
Thus, our proposed truncation scheme in Eq.~\eqref{ch4-TS} successfully closes the hierarchy and makes the calculation of $\mathcal{C}^{QQ}_{r}(t',t)$ possible.

For calculational convenience, we now represent the correlation functions in the Fourier space by using the following transformation,
\begin{eqnarray}\label{ch4-Fourier_transform}
\tilde{\mathcal{C}}_{n}^{AB}(t',t)=\sum_{r=0}^{L-1}\mathcal{C}_{r}^{AB}(t',t) e^{iq_nr},
\end{eqnarray} 
where the inverse Fourier transform is given by
\begin{eqnarray}
\label{ch4-Inverse_Fourier_transform}
\mathcal{C}_{r}^{AB}(t',t)=\frac{1}{L}\sum_{n=0}^{L-1}\tilde{\mathcal{C}}_{n}^{AB}(t',t) e^{-iq_nr},
\end{eqnarray}
with 
\begin{eqnarray}
q_n=\frac{2\pi n}{L}.
\end{eqnarray}
Indeed, using the Fourier transform Eq.~\eqref{ch4-Inverse_Fourier_transform} in Eqs.~\eqref{ch4-time-evolution-Q-Q-general_2} and \eqref{ch4-time-evolution-eta-Q-diff_time}, we obtain much simplified forms of the time-evolution equations for the respective correlation functions $\mathcal{\tilde{C}}^{QQ}_{n}(t',t)$ and $\mathcal{\tilde{C}}^{\eta Q}_{n}(t'',t)$ in the Fourier space. We write the corresponding solutions as follows:
\begin{eqnarray}\label{ch4-Q-Q_solution1}
\mathcal{\tilde{C}}^{QQ}_{n}(t',t) &=& D(\rho) \hspace{-0.1 cm} \int_{t}^{t'} \hspace{-0.25 cm} dt'' \left(1-e^{-i q_n} \right) \mathcal{\tilde{C}}^{\eta Q}_{n}(t'',t) + \mathcal{\tilde{C}}^{QQ}_{n}(t,t), \nonumber \\ \\
\label{ch4-eta-Q_solution1}
\mathcal{\tilde{C}}^{\eta Q}_{n}(t'',t) &=& e^{-\lambda_{n} D(\rho) (t''-t)} \mathcal{\tilde{C}}^{\eta Q}_{n}(t,t), 
\end{eqnarray}
where $t' \geq t'' \geq t$; the quantity $\lambda_{n}$ is an eigenvalue of the negative discrete Laplacian operator and is given by 
\begin{eqnarray}\label{ch4-eigen-value}
\lambda_{n}=2\left(1- \cos q_n \right). 
\end{eqnarray}
Eqs.~\eqref{ch4-Q-Q_solution1} and \eqref{ch4-eta-Q_solution1} clearly suggest that unequal-time correlation functions $\mathcal{\tilde{C}}^{QQ}_{n}(t',t)$ and $\mathcal{\tilde{C}}^{\eta Q}_{n}(t'',t)$ solely depend on their equal-time counterparts. Thus, to calculate the equal-time density-current correlation function $\mathcal{\tilde{C}}^{\eta Q}$, we first derive its infinitesimal time-evolution equation in real space. Then performing Fourier transformation, as defined in Eq.~\eqref{ch4-Fourier_transform}, we obtain the time-evolution equation for $\mathcal{\tilde{C}}^{\eta Q}_{n}(t,t)$, and the corresponding solution is given by
\begin{eqnarray}\label{ch4-eta1-Q2_sametime}
\mathcal{\tilde{C}}^{\eta Q}_{n}(t,t)=\int_{0}^{t}dt''' e^{-\lambda_{n} D(\rho)(t-t''')}\mathcal{\tilde{S}}^{\eta Q}_{n}(t'''),
\end{eqnarray}
where the term $\mathcal{\tilde{S}}^{\eta Q}_{q}(t''')$ takes the following form:
\begin{eqnarray}\label{ch4-source-eta-Q}
\mathcal{\tilde{S}}^{\eta Q}_{n}(t) &=& \frac{1}{(1-e^{-iq_{n}})}\left[D(\rho) \lambda_{n} \mathcal{\tilde{C}}^{\eta \eta}_{n}(t,t) - {\mathcal{F}_{n}(t)} \right].~~~
\end{eqnarray}
Notably, the quantity ${\cal F}_{n}(t)$ is directly related to gap distribution  $P(g, t)$ and is given by
\begin{eqnarray}\label{ch4-f_{q}}
\mathcal{F}_{n}(t)&=&\rho \Big[\alpha \lambda_{n} \Big\{1 - P(0,t)\Big\} + \beta \Big\{\sum_{g=1}^{l_0-1}  \lambda_{g n}P(g, t) +  \lambda_{l_0 n}\sum_{g=l_0}^{\infty}P(g, t)\Big\}\Big],
\end{eqnarray}
and $\mathcal{\tilde{C}}^{\eta \eta}_{n}(t,t)$ is the equal-time density-density correlation function, which, in the steady state, can be simply written as 
\begin{eqnarray}\label{ch4-static-density-correlation}
\mathcal{\tilde{C}}^{\eta \eta}_{n}=\frac{\mathcal{F}_{n}}{2D(\rho)\lambda_{n} }.
\end{eqnarray} 
Here, the steady state (i.e., time-independent) $\mathcal{F}_{n}$ is obtained by substituting the gap distribution  $P(g)$, which is obtained in the steady state and therefore is independent of time. 
Now, by replacing the steady-state quantities $\mathcal{\tilde{C}}^{\eta \eta}_{n}$ and $\mathcal{F}_{n}$ in Eq.~\eqref{ch4-source-eta-Q} and then using Eqs.~\eqref{ch4-eta1-Q2_sametime} and \eqref{ch4-eta-Q_solution1}, we obtain the steady-state unequal-time correlation for the density and current,
\begin{eqnarray}
\mathcal{\tilde{C}}^{\eta Q}_{n}(t'',t)=\frac{-\mathcal{F}_{n}e^{-\lambda_{n} D(\rho)t''}}{2D(\rho)\lambda_{n}(1-e^{-iq_n})}\left(e^{\lambda_{n} D(\rho)t} - 1\right).
\label{ch4-density_current_unequal_soln}
\end{eqnarray}

Now, using Eq.~\eqref{ch4-density_current_unequal_soln} in Eq.~\eqref{ch4-Q-Q_solution1} and following the formalism of Ref.~\cite{Chakraborty_PRE_2024_RTP_II}, we first calculate the two-point equal-time correlation function for time-integrated bond current and finally obtain the dynamic current-current correlation function for the time-integrated bond currents at two space and time points,
	\begin{eqnarray}
    \label{ch4-Q1_Q2_solution_most_general}
 \hspace{-0 cm}\mathcal{C}^{QQ}_{r}(t',t) &=& \Gamma_{r} t -\frac{1}{2LD} \sum_{n=1}^{L-1}\frac{\mathcal{F}_{n}}{\lambda_n^{2}}\left(e^{-\lambda_{n}Dt}- e^{-\lambda_{n}Dt'}\right)\left(e^{-\lambda_{n}Dt}- 1\right) e^{-iq_nr} \nonumber \\ && - \frac{1}{2L}\sum_{n=1}^{L-1}\frac{\mathcal{F}_{n}}{\lambda_n}\Bigg\{t - \frac{\left(1 - e^{-\lambda_{n}Dt} \right)}{\lambda_{n}D}  \Bigg\}(2-\lambda_{rn}).
\end{eqnarray}
It is important to note that the quantity $\Gamma_{r}$ actually corresponds to the strength of the spatiotemporal correlation of the fluctuating component $J^{fl}_{r}$, i.e., $\langle J_{r}^{fl}(t) J_{0}^{fl}(0) \rangle= \Gamma_{r} \delta(t)$, which is derived later in Eq. \eqref{ch4-fluc_current_correlation}.

\subsection{Space-time-integrated current fluctuations}
\label{Sec:ch4-total_current_fluc}

So far, in our analysis, we have restricted ourselves to calculating the variance of time-integrated current across a single bond. In this section, we study the fluctuation (i.e., the variance or the second cumulant) characteristics of the space-time-integrated current $Q_{tot}(L, T)$ across the entire system,
\begin{eqnarray}
\label{ch4-space-time-integrated-current}
Q_{tot}(L,T)= \sum_{i=0}^{L-1}Q_{i}(T) = \int_{0}^{T}dt\sum_{i=0}^{L-1}J_{i}(t),
\end{eqnarray}
where, in the steady state, the average current $\langle Q_{tot}(L,T) \rangle =0$ vanishes on a periodic domain.
Now, utilizing the decomposition scheme in Eq.~\eqref{ch4-current_decompose}, we can break down the instantaneous bond current $J_{i}(t)$ into its diffusive component $J^{D}_{i}(t)$ and fluctuating component $J^{(fl)}_{i}(t)$. Additionally, since the system under consideration here is periodic, we can further utilize the identity $\sum_{i=0}^{L-1}J^{D}_{i}(t)=0$. This immediately enables us to express the space-time-integrated current as
\begin{eqnarray}\label{ch4-space-time-integrated-current_2}
Q_{tot}(L,T)= \int_{0}^{T}dt\sum_{i=0}^{L-1}J^{fl}_{i}(t),
\end{eqnarray}
and therefore we have the variance expressed in terms of fluctuating current correlations,
\begin{eqnarray}\label{ch4-total_current_fluc}
\hspace{-2.5cm}
\langle Q^{2}_{tot}(L,T) \rangle_c = \langle Q^{2}_{tot}(L,T) \rangle - \langle Q_{tot}(L,T) \rangle^2 &=& \int_{0}^{T}dt_1 \int_{0}^{T}dt_2\sum_{i=0}^{L-1} \sum_{r}\langle  J^{fl}_{i+r}(t_1) J^{fl}_{i}(t_2)\rangle_c \nonumber \\ &=&  LT \int_{0}^{T}dt \sum_{r} \mathcal{C}^{J^{fl}J^{fl}}_{r}(t,0).
\end{eqnarray}
Note that, as $\langle Q_{tot}(L,T) \rangle=0$ in the steady state, we have $\langle Q^{2}_{tot}(L,T) \rangle_c = \langle Q^{2}_{tot}(L,T) \rangle$, the second moment of the space-time-integrated current.
Now, to characterize the total current fluctuation across the entire system, we must first calculate the space-time correlation of the fluctuating current $\mathcal{C}^{J^{fl}J^{fl}}_{r}(t, 0)= \langle J^{fl}_{r}(t) J^{fl}_{0}(0) \rangle_c$, which, using Eq.~\eqref{ch4-current_decompose} can be expressed in terms of the following correlation functions:
\begin{eqnarray}\label{ch4-inst_current_correlation_decomposition}
\mathcal{C}^{J^{fl}J^{fl}}_{r}(t, 0)=\mathcal{C}^{JJ}_{r}(t, 0) - \mathcal{C}^{J^{D}J}_{r}(t, 0) - \mathcal{C}^{J^{fl}J^{D}}_{r}(t, 0). 
\end{eqnarray}
Here, we have used the fact that fluctuation current $J^{(fl)}(t)$ at time $t$ is not correlated with the diffusive current $J^{D}_{0}(0)$ at an earlier time $t=0$, i.e., we have
\begin{eqnarray}\label{ch4-fluc_diffusive_curr_correlation}
\mathcal{C}^{J^{fl}J^{D}}_{r}(t, 0) = \langle J^{fl}_{r}(t) J^{D}_{0}(0) \rangle =0.
\end{eqnarray}
Then the third term in Eq.~\eqref{ch4-inst_current_correlation_decomposition} immediately drops out. To solve for the first term, i.e., the two-point instantaneous current correlation function $\mathcal{C}^{JJ}_{r}(t, 0)$, we use the following formula:
\begin{eqnarray}\label{ch4-inst_current_correlation_formula}
\mathcal{C}^{JJ}_{r}(t,0)=\frac{d}{dt}\frac{d}{dt''}\left[\mathcal{C}^{QQ}_{r}(t,t'') \Theta(t-t'') \right]_{t''=0}.
\end{eqnarray}
Upon plugging in $\mathcal{C}^{QQ}_{r}(t,t'')$ from Eq.~\eqref{ch4-Q1_Q2_solution_most_general}, the above equation simply yields the following expression:
\begin{eqnarray}\label{ch4-inst_current_correlation_formula-1}
\mathcal{C}^{JJ}_{r}(t,0)=\Gamma_{r} \delta(t) - \frac{D}{4L}\sum_{n}\left(2 - \lambda_{rn} \right) \mathcal{F}_{n} e^{-\lambda_{n}Dt}.
\end{eqnarray}
Moreover, in order to determine the second term $\mathcal{C}^{J^{D}J}_{r}(t, 0)$, we use the following relation:
\begin{eqnarray}\label{ch4-diff_inst_correlation_1}
\mathcal{C}^{J^{D}J}_{r}(t, 0) &=&\left[\frac{d}{dt''}\mathcal{C}^{J^{D}Q}_{r}(t, t'')\right]_{t''=0},  \\ &\simeq& D\frac{d}{dt''} \left[\mathcal{C}^{\eta Q}_{r}(t, t'')-\mathcal{C}^{\eta Q}_{r+1}(t, t'') \right]_{t''=0},\label{ch4-diff_inst_correlation_2}
\end{eqnarray}
where we use the closure approximation as in Eq.~\eqref{ch4-closure_approximation} and arrive  at Eq.~\eqref{ch4-diff_inst_correlation_2} by using  Eq.~\eqref{ch4-diff_inst_correlation_1}. Furthermore, following Eq.~\eqref{ch4-Fourier_transform}, we expand the correlation function  $\mathcal{C}^{\eta Q}_{r}(t, t')$ in the Fourier basis and, then using Eq.~\eqref{ch4-density_current_unequal_soln}, we obtain the desired solution,
\begin{eqnarray}\label{ch4-diff_inst_correlation_solution}
\mathcal{C}^{J^{D}J}_{r}(t, 0)=-\frac{D}{4L}\sum_{n}(2-\lambda_{rn})\mathcal{F}_{n}(t)e^{-\lambda_n Dt}.
\end{eqnarray}
Importantly, the above solution coincides with the two-point unequal-time correlation, $\mathcal{C}^{JJ}_{r}(t, 0)$, which is displayed in the second term of Eq.~\eqref{ch4-inst_current_correlation_formula-1} with $t \geq t'=0$. Finally, using Eqs.~\eqref{ch4-inst_current_correlation_formula-1}, \eqref{ch4-fluc_diffusive_curr_correlation} and \eqref{ch4-diff_inst_correlation_solution} in Eq.~\eqref{ch4-inst_current_correlation_decomposition}, we readily obtain
\begin{eqnarray}
\label{ch4-fluc_current_correlation}
\mathcal{C}^{J^{fl}J^{fl}}_{r}(t, 0)= \langle J^{fl}_{r}(t) J^{fl}_{0}(0) \rangle= \delta(t) \Gamma_{r}.
\end{eqnarray}
That is, the fluctuating current components are delta-correlated in time but have finite correlation $\Gamma_r$ in space, as given in Eq.~\eqref{ch4-gamma_r}. Finally, using Eq.~\eqref{ch4-fluc_current_correlation} in Eq.~\eqref{ch4-total_current_fluc}, we find that the steady-state variance of space-time-integrated current across the entire system satisfies the following relation:
\begin{eqnarray}\label{ch4-I_fluctuation1}
\frac{1}{LT} \langle Q^{2}_{tot}(L,T)\rangle_c = \frac{1}{LT}\langle Q^{2}_{tot}(L,T)\rangle = 2 \chi(\rho),
\end{eqnarray}
where we employ the identity,
\begin{eqnarray}
\sum_{r} \Gamma_{r}(\rho) = \rho \left[ \alpha \left(1-P(0)\right) + \beta \left(\sum_{g=1}^{l_0-1} g^{2}P(g) + l_0^{2}\sum_{g=l_0}^{\infty} P(g)\right) \right] =  2\chi(\rho),
\label{ch4-identity_gamma}
\end{eqnarray}
with $\chi(\rho)$ being the collective particle mobility. We explicitly verify Eqs.~\eqref{ch4-I_fluctuation1} and \eqref{ch4-identity_gamma} with our numerical simulation data in sections \ref{sec:ch-4_result_finite_range} and \ref{sec:ch-4_result_infinite_range}.

\subsection{Steady-state fluctuation of time-integrated bond current}
\label{Sec:ch4-bond_current_fluc}

We immediately obtain the steady-state variance of time-integrated bond current, $\mathcal{C}^{QQ}_{0}(T, T) \equiv \langle Q^{2}(T) \rangle_c$, by putting $r=0$ and $t'=t=T$ in Eq.~\eqref{ch4-Q1_Q2_solution_most_general}, and the resulting expression is given by,
\begin{eqnarray}\label{ch4-bond-current-fluc-0}
\mathcal{C}^{QQ}_{0}(T,T) = \Gamma_{0} T - \frac{1}{L} \sum_{n}\frac{\mathcal{F}_n}{\lambda_n} \Bigg\{T - \frac{\left(1 - e^{-\lambda_{n} DT} \right)}{\lambda_{n} D}  \Bigg\},
\end{eqnarray}
which, upon using the identity
\begin{eqnarray}
\sum_{n}\left(\frac{\mathcal{F}_n}{\lambda_n}\right) = L \Gamma_{0} - 2 \chi(\rho,\gamma),
\end{eqnarray} 
can be immediately written in the following form:
\begin{eqnarray}\label{ch4-bond-current-fluc}
\mathcal{C}^{QQ}_{0}(T,T) = \frac{2 \chi(\rho)}{L} T + \frac{1}{D(\rho)L}\sum_{n}\frac{\mathcal{F}_n}{\lambda_n^{2}} \left(1 - e^{-\lambda_{n} D(\rho)T} \right). \nonumber \\
\end{eqnarray}
Notably, $\mathcal{C}^{QQ}_{0}(T, T) =\langle Q^{2}(T) \rangle_c  = \langle Q^{2}(T) \rangle$ in Eqs.~\eqref{ch4-bond-current-fluc-0} and \eqref{ch4-bond-current-fluc} share the identical form with the one derived for the long-ranged lattice gas (LLG) in \cite{Chakraborty_PRE_2024_RTP_II}. Therefore, to obtain the limiting behaviors of $\langle Q^{2}(T) \rangle$ for LLG, we apply the same technique employed for LLG in \cite{Chakraborty_PRE_2024_RTP_II}, which immediately yields the following behaviors:
\begin{eqnarray} 
 \langle Q^{2}(T) \rangle \simeq 
\left\{
\begin{array}{ll}
\vspace{0.15 cm}
  \Gamma_{0} T,            ~~~  & {\rm for}~   DT \ll 1, \\
\vspace{0.15 cm}
  \frac{2\chi(\rho)}{\sqrt{\pi D(\rho)}}\sqrt{T},            ~~~  & {\rm for}~  1 \ll DT \ll L^2, \\
\vspace{0.15 cm}
 \frac{2\chi(\rho)}{L} T,            ~~~  & {\rm for}~ DT \gg L^{2} \\
\end{array}
\right.
\label{ch4-cf_limit}
\end{eqnarray}
The above equation indicates the existence of three distinct time regimes: (a) in the initial regime, $T \ll 1/D$, $\langle Q^{2}(T) \rangle$ follows linear growth, and the prefactor $\Gamma_0$ is readily determined from Eq.~\eqref{ch4-gamma_r} by setting $r=0$, (b) within the intermediate regime, $1/D \ll T \ll L^{2}/D$, $\langle Q^{2}(T) \rangle$ exhibits subdiffusive growth, with the prefactor dependent on the mobility $\chi(\rho)$ and the bulk-diffusion coefficient $D(\rho)$, (c) in the final regime, $T \gg L^{2}/D$, the subdiffusive growth transitions to a linear or diffusive one, where the growth law is solely governed by the mobility $\chi(\rho)$ and the system size $L$.

Interestingly, at large times $T \gg 1/D$ and large system sizes, the rhs of Eq.~\eqref{ch4-bond-current-fluc} contributes in \textcolor{black}{the small wave-vector magnitude} $q \equiv q_n  \rightarrow 0$ limit, where, for brevity, we simply denote $q_n$ as $q$. Therefore, to characterize fluctuation at large times, we need to perform a small $q$ expansion in Eq.~\eqref{ch4-bond-current-fluc}. This is done by simply putting $\lambda_q \simeq q^{2}$, which quite strikingly yields $\textcolor{black}{{\cal F}_q}/\lambda_q = \chi$ and the resulting equation can be expressed as the following scaling form:
\begin{eqnarray}\label{ch4-bond-current-fluc-3}
 \frac{D}{2\chi L} \langle Q^{2}(T) \rangle = \mathcal{W}\left(\frac{DT}{L^{2}}\right).
\end{eqnarray} 
The scaling function is calculated exactly within the truncation scheme and is given by the following series,
\begin{eqnarray}\label{ch4-cf_scaling_function}
\mathcal{W}\left(y\right) =y + \lim_{L \rightarrow \infty} \frac{1}{L^{2}}\sum_{n}\frac{1}{\lambda_n}\left(1 - e^{-\lambda_{n}yL^{2}} \right).
\end{eqnarray} 
Upon performing an asymptotic analysis to obtain the behavior of $\mathcal{W}\left(y\right)$ in the two limiting cases when $y \ll 1$ and $y \gg 1$, we obtain
\begin{eqnarray} 
 \mathcal{W}\left(y\right) \simeq 
\left\{
\begin{array}{ll}
\vspace{0.15 cm}
  \sqrt{y/\pi},            ~~~  & {\rm for}~ y \ll 1 , \\
\vspace{0.15 cm}
 y,            ~~~  & {\rm for}~ y \gg 1. \\
\end{array}
\right.
\label{ch4-cf_scaling_function_limit}
\end{eqnarray}
Undoubtedly, Eqs.~\eqref{ch4-cf_limit}, \eqref{ch4-bond-current-fluc-3}, \eqref{ch4-cf_scaling_function}, and \eqref{ch4-cf_scaling_function_limit} constitute the main results in our analysis so far. In sections \ref{sec:ch-4_result_finite_range} and \ref{sec:ch-4_result_infinite_range}, respectively, we verify the results for the system having finite-range hopping ($l_0=2$) as well as that having infinite-range hopping both at and away from the critical point.

\section{Special cases}
\label{sec:ch4-results}

In the previous section, using a microscopic dynamical framework, we have calculated current fluctuations in terms of two macroscopic transport coefficients - the bulk-diffusion coefficient $D(\rho)$ and the mobility $\chi(\rho)$. In this section, we study the two analytically tractable cases, $l_0=2$ and $l_0 \rightarrow \infty$, and obtain explicit closed-form expressions for the two transport coefficients, the bulk-diffusion coefficient and the mobility, as a function of density.
For simplicity, from now on, we consider $\alpha=\beta=1/2$.

\subsection{Case I: Finite-ranged hopping (FRH) with \texorpdfstring{$l_0=2$}{l0=2} }
\label{sec:ch-4_result_finite_range}

\begin{figure*}[tpb]
           \centering
         \includegraphics[width=0.495\linewidth]{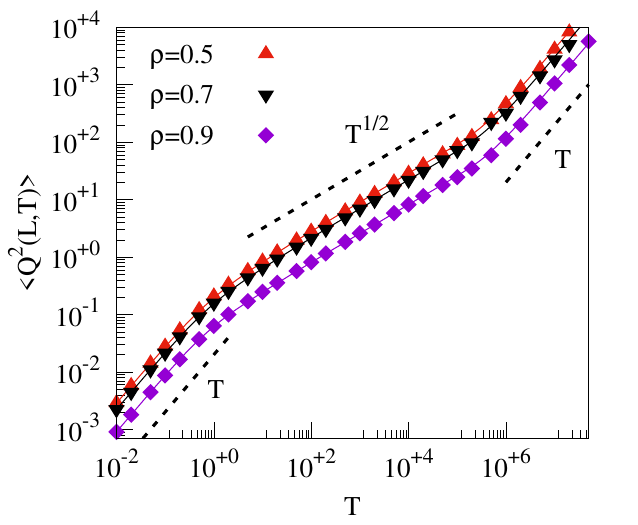}\hfill
         \includegraphics[width=0.495\linewidth]{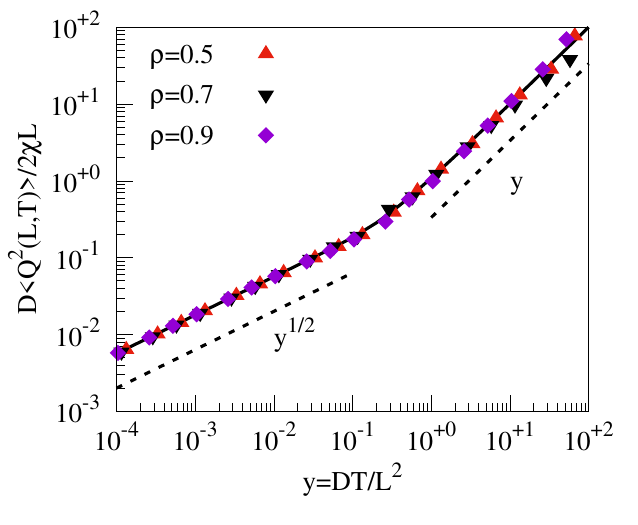}
         \caption{\small{{\it Finite-ranged hopping with $l_0=2$.} Left panel: We plot the steady-state variance of time-integrated bond-current  $\langle Q^{2}(L, T) \rangle_c = \langle Q^{2}(L, T) \rangle$ as a function of time $T$, where the data are obtained from simulations (points) for $\rho=0.5,$ $0.7$, and $0.9$ at $L=1000$. We then compare the simulation results with the analytical ones as in Eq.~\eqref{ch4-bond-current-fluc} (line). The variance of time-integrated bond current exhibits diffusive growth at early times, subdiffusive growth at the intermediate times, and finally diffusive (linear) growth at large times, as shown by the dotted lines, which are consistent with Eq.~\eqref{ch4-cf_limit}. Right panel: The scaled bond-current fluctuation $D_2\langle Q^{2}(L, T) \rangle/2\chi_2 L$ is plotted against the rescaled hydrodynamic time $y=D_2(\rho)T/L^{2}$ for the above-mentioned combination of densities and system size. We also compare the scaling collapse obtained from simulations with the analytic solution of the scaling function $\mathcal{W}\left(y\right)$ as in Eq.~\eqref{ch4-cf_scaling_function} (red line).}}
          \label{fig:ch4_bcf_l2}
 \end{figure*}
 
In the case of finite-ranged hopping, it is possible to analytically obtain the transport coefficients explicitly as a function of density. First, we note that the bulk-diffusion coefficient $D(\rho)$ and the mobility $\chi(\rho)$ can be expressed in terms of the steady-state gap distribution $P(g)$, as derived in Eqs.~\eqref{ch4-bulk-diffusivity-LH} and \eqref{ch4-identity_gamma}, respectively.  Indeed, the direct numerical determination of the two transport coefficients is possible for an arbitrary density $\rho$ and hopping range $l_0$. However, as discussed below, analytically calculating the transport coefficients is presently limited to the simplest, albeit nontrivial, case of the hopping range $l_0=2$. In this section, we thus restrict ourselves to solving for this particular limit and verifying our analytical findings of current fluctuations with Monte Carlo simulations. 
\begin{figure}[tpb]
           \centering
         \includegraphics[width=0.75\linewidth]{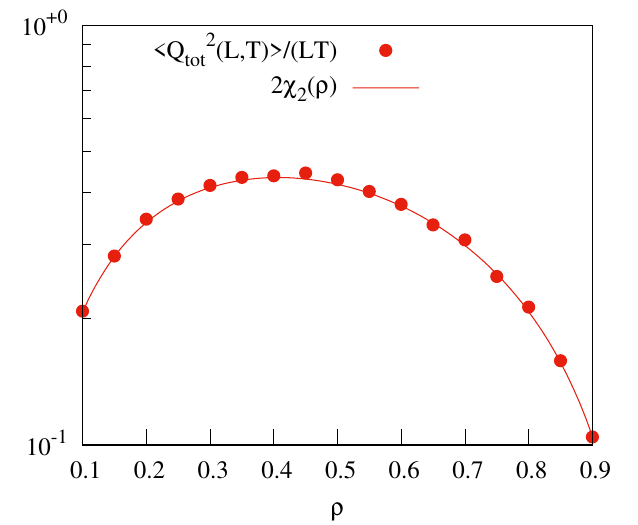}
         \caption{\small{\textit{Verification of Eq.~\eqref{ch4-I_fluctuation1} for $l_0=2$:} We plot the numerically obtained scaled variance $\langle Q^{2}_{tot}(L,T)\rangle_c/LT = \langle Q^{2}_{tot}(L,T)\rangle / LT$ of the space-time-integrated current across the entire system as a function of $\rho$ and compare the numerical data with the theoretically calculated mobility $\chi_{2}(\rho)$ (line), as in Eq.~\eqref{ch4-chi_l2_final}.}}
          \label{fig:ch4_tcf_l2}
 \end{figure}
To begin with, we first note that this exclusion model can be directly mapped to an equivalent dual model with an unbounded mass transfer model studied in \cite{subhadip_PRE_2021} where the gap distribution function in the former is identical to the mass distribution of the latter.. In this specific limit of $l_0=2$ with $\alpha = \beta = 1/2$, the mass distribution of the unbounded model has been calculated to be the following \cite{subhadip_PRE_2021}:
\begin{equation}
\label{pg_l2}
P(g)=\left( \frac{P_1}{P_0} \right)^g P_0 F_{g+1},
\end{equation}
where $F_{g+1}$ is the $(g+1)$th element of the Fibonacci series \cite{beck2010art} and the prefactors $P_0$ and $P_1$ both depend on the mass density $\tilde{\rho}= (1/\rho -1)$ as follows:
\begin{eqnarray}
\label{P0}
P_0 = \frac{9+5\tilde{\rho}-\sqrt{1+10\tilde{\rho}+5\tilde{\rho}^{2}}}{2(2+\tilde{\rho})^{2}}, ~~~~~~~~~~~~~~~~~~~~~~ \\
\label{P1}
P_1 = \frac{(3\tilde{\rho}+5)\sqrt{1+10\tilde{\rho}+5\tilde{\rho}^{2}}-(5\tilde{\rho}^{2}+12\tilde{\rho}+5)}{2(2+\tilde{\rho})^{3}}.
\end{eqnarray}
Finally, upon using the above $P(g)$ in Eqs.~\eqref{ch4-bulk-diffusivity-LH} and \eqref{ch4-identity_gamma} and after some algebraic manipulations, the analytic expressions of the bulk-diffusion coefficient $D(\rho) \equiv D_2(\rho)$ and the mobility $\chi(\rho) \equiv \chi_2(\rho)$ are obtained  explicitly as a function of density,
\begin{eqnarray}\label{ch4-D_l2_final}
\hspace{-1.5 cm} D_2(\rho)&=&\frac{1}{8(1+\rho)^{4}}  \Bigg[10\Big(1-\frac{\rho}{\sqrt{5-4\rho^{2}}}\Big) + 5\rho \Big(6 + \frac{\rho}{\sqrt{5-4\rho^{2}}}\Big) + \rho^{2} \Big(17 + \frac{12\rho}{\sqrt{5-4\rho^{2}}}\Big) \Bigg],~~~~~  \\
  \hspace{-1.5 cm}\chi_2(\rho)&=&\frac{\rho}{8(1+\rho)^{3}}  \Bigg[10 + 4 \rho \left( 5 - \sqrt{5-4\rho^{2}} \right) - \rho^{2} \Big(9 + \sqrt{5-4\rho^{2}}\Big) - 16 \rho^{3}\Bigg].
\label{ch4-chi_l2_final}
\end{eqnarray} 
It is important to highlight that both $D_2(\rho)$ and $\chi_2(\rho)$ remain finite and bounded within the entire density range $0 < \rho < 1$. Now we proceed to validate the above theoretical predictions in the subsequent Monte Carlo analysis of the model.

We first verify the analytic results for the variance of time-integrated bond current $\langle Q^{2}(T) \rangle_{c}$, as derived in Eqs.~\eqref{ch4-bond-current-fluc} and \eqref{ch4-cf_limit}, for $l_0=2$. In Fig.~\ref{fig:ch4_bcf_l2} (left-panel) we plot the numerically obtained $\langle Q^{2}(T) \rangle_{c}$ as a function of the observation time $T$ for various densities $\rho=0.5$, $0.7$, and $0.9$. We observe that the simulation data points initially (for small times) display diffusive or linear $\sim T$ growth, which then crosses over to a subdiffusive $\sim \sqrt{T}$ growth and eventually to diffusive or linear $\sim T$ growth, as shown in Eq.~\eqref{ch4-cf_limit}. We also complement the simulation data points with the theoretical lines obtained from Eq.~\eqref{ch4-bond-current-fluc}, where we substitute the transport coefficients from Eqs.~\eqref{ch4-D_l2_final} and \eqref{ch4-chi_l2_final}, and solve for $\mathcal{F}_{n}$ utilizing the steady-state gap distribution $P(g)$ obtained from Eq.~\eqref{pg_l2} in Eq.~\eqref{ch4-f_{q}}. Indeed, we observe a remarkable agreement between the simulation data (points) and theoretically obtained results (lines). These findings overall verify Eqs.~\eqref{ch4-bond-current-fluc} and \eqref{ch4-cf_limit} for lattice gases with finite-ranged hopping $l_0=2$.

 To verify the scaling form given in Eq.~\eqref{ch4-bond-current-fluc-3}, we next plot the numerically obtained scaled variance of time-integrated bond current, $D_2 \langle Q^{2}(T) \rangle/2\chi_2 L$, as a function of the scaled (hydrodynamic) time $y=D_2 T/L^{2}$ in Fig.~\ref{fig:ch4_bcf_l2}(right panel); here we consider the following set of densities $-$ $\rho=0.5$, $0.7$, and $0.9$. We find the simulation data points collapse onto each other remarkably well, and the scaling collapse obtained from the simulation data points is in excellent agreement with the theoretically derived scaling function $\mathcal{W}(y)$ (red solid line) given in Eq.~\eqref{ch4-cf_scaling_function}. Moreover, we observe the scaling function as well as the collapsed data points make a crossover from the $\sqrt{y}$ growth at small times to the linear or $y$ behavior at very large times, thus verifying Eqs.~\eqref{ch4-bond-current-fluc-3}, \eqref{ch4-cf_scaling_function}, and \eqref{ch4-cf_scaling_function_limit}.

We also verify the equilibrium-like fluctuation-response relation in Eq.~\eqref{ch4-I_fluctuation1} by numerically computing the variance of space-time-integrated current $\langle Q^{2}_{tot}(L, T)\rangle_{c}$ in the steady state. We calculate $\langle Q^{2}_{tot}(L, T)\rangle_{c}$ numerically for system size $L=1000$ and observation time $T=100$. To verify Eq.~\eqref{ch4-I_fluctuation1}, we now plot the numerically obtained scaled variance $\langle Q_{tot}^{2}(L, T) \rangle/2LT$ (close points), as a function of density $\rho$, in Fig.~\ref{fig:ch4_tcf_l2}. We compare the simulation findings with the analytical solution $2\chi_{2}(\rho)$ obtained in Eq. \eqref{ch4-chi_l2_final}, which is represented as lines. We again observe a nice agreement between the simulation data points and theoretical results, thus confirming Eq.~\eqref{ch4-I_fluctuation1}.
\begin{figure}[tpb]
           \centering
         \includegraphics[width=0.75\linewidth]{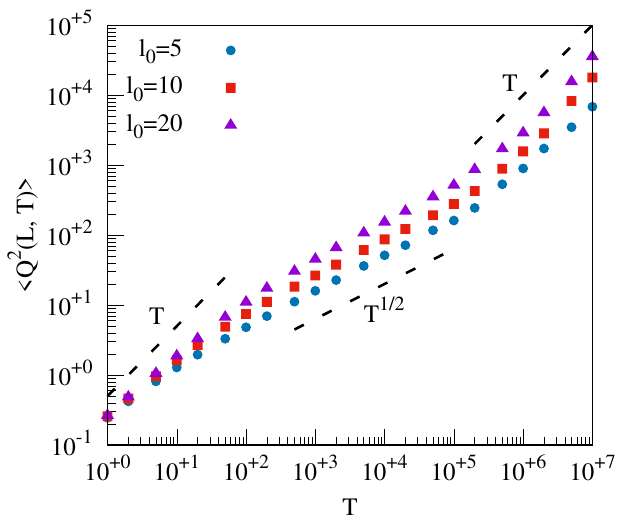}
         \caption{\small{The steady-state variance $\langle Q^{2}(L, T) \rangle_c = \langle Q^{2}(L, T) \rangle$ of time-integrated bond current  is plotted against the observation time $T$ for various hopping range $l_0=5$, $10$, and $20$ at density $\rho=0.5$ and system size $L=1000$. The dotted lines represent our analytical predictions for the growth law in three time regimes as shown in Eq.~\eqref{ch4-cf_limit}.}}
          \label{fig:ch4_tcf_l_various}
 \end{figure}

\begin{figure*}[tpb]
           \centering
         \includegraphics[width=0.495\linewidth]{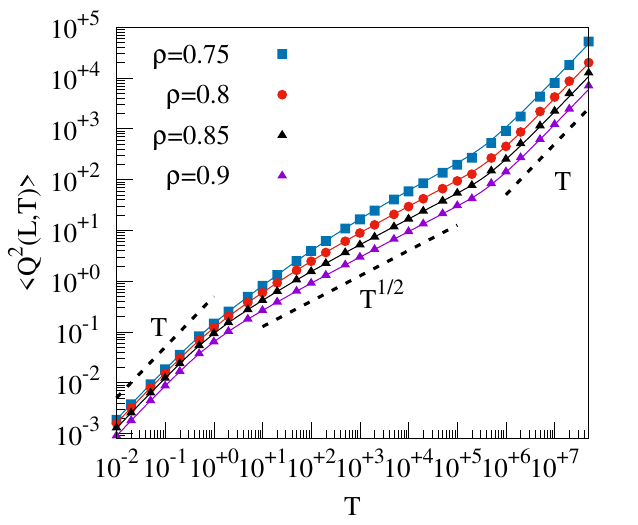}\hfill
         \includegraphics[width=0.495\linewidth]{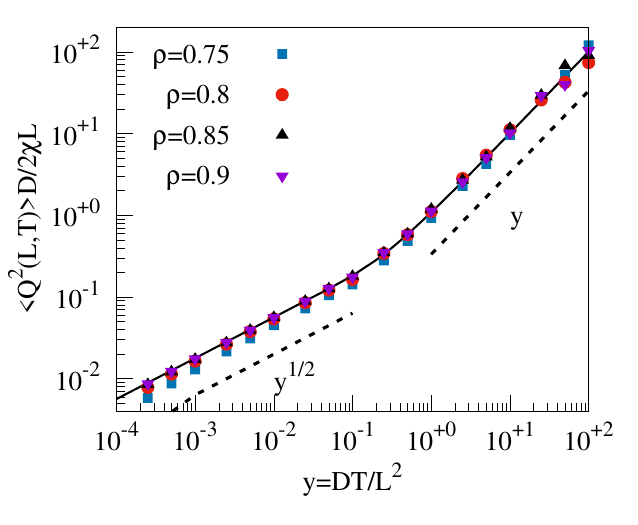}
         \caption{\small{\textit{Current fluctuations for infinite-ranged hopping.} Left: We plot the steady-state variance $\langle Q^{2}(L, T) \rangle_c = \langle Q^{2}(L, T) \rangle$ of time-integrated bond-current as a function of time $T$, obtained from simulations (points) for $l_0 \rightarrow \infty$ in disordered phase ($\rho > \rho_c$) with densities $\rho=0.75,$ $0.8$, $0.85$ and $0.9$. We also compare simulations with the theoretically obtained solution Eq.~\eqref{ch4-bond-current-fluc} (line). The variance $\langle Q^{2}(L, T) \rangle $ exhibits diffusive growth at early times, subdiffusive growth at the intermediate times, and a diffusive (linear) growth at large times, as shown by the dotted lines. Right: The scaled bond-current fluctuation $D_{\infty}\langle Q^{2}(L, T) \rangle/2\chi_{\infty} L$ is plotted against the rescaled time $y=D_{\infty}(\rho)T/L^{2}$ for the above-mentioned densities. We also compare the numerically obtained scaling collapse with the analytic scaling function $\mathcal{W}\left(y\right)$ as in Eq.~\eqref{ch4-cf_scaling_function} (red line).}}
          \label{fig:ch4_bcf_linf}
 \end{figure*}
 \begin{figure}[tpb]
           \centering
         \includegraphics[width=0.75\linewidth]{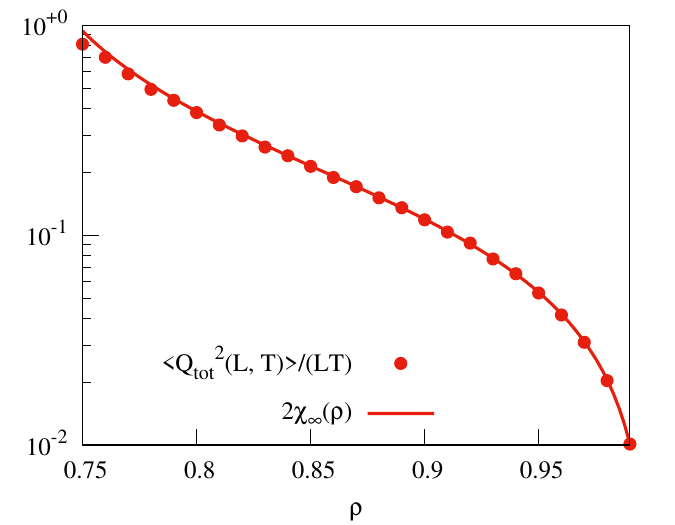}
         \caption{\small{\textit{Verification of Eq.~\eqref{ch4-I_fluctuation1} for $l_0 \rightarrow \infty$ in disordered phase (i.e., $\rho > \rho_c$):} We plot the numerically obtained scaled fluctuation of the space-time integrated current as a function of $\rho$ in the disordered phase and compare the numerical data with the theoretically calculated $\chi_{\infty}(\rho)$ (solid line), as derived in Eq.~\eqref{ch4-chi-inf-final}.}}
          \label{fig:ch4_tcf_linf}
 \end{figure}
  Indeed, as already mentioned, our theory should remain valid for arbitrary $l_0$; though the explicit formula for current fluctuations and the Onsager transport coefficients could not be obtained in the present work. However, the qualitative behavior of dynamic fluctuation properties remains, perhaps not surprisingly, the same as that in the case of $l_0=2$, and we observe a similar growth behavior for the variance of time-integrated bond current for intermediate values of \( l_0 \); of course, there are quantitative differences as the current fluctuations now tend to grow with increasing $l_0$. To demonstrate this, in Fig.~\ref{fig:ch4_tcf_l_various}, we plot the variance of the time-integrated bond current, \( \langle Q^{2}(L, T) \rangle \), as a function of the observation time \( T \) for \( l_0 = 5 \), \( 10 \), and \( 20 \), while keeping the density fixed at \( \rho = 0.5 \) and the system size at \( L = 1000 \). We find that, although the magnitude of the variance increases with \( l_0 \), the overall growth trend remains consistent with the asymptotic predictions as given in Eq.~\eqref{ch4-cf_limit}.

\subsection{Case II: Infinite-range hopping (IRH) with $l_0 \rightarrow \infty$}\label{sec:ch-4_result_infinite_range}

However, as discussed below, for infinite-ranged hopping, the situation qualitatively changes due to the anomalous current fluctuations present in the system.
To demonstrate this, we first set $l_0 \rightarrow \infty$ and $\alpha = \beta = 1/2$ in Eqs.~\eqref{ch4-bulk-diffusivity-LH} and \eqref{ch4-identity_gamma}. We find that both the bulk-diffusion coefficient and the total current fluctuation strength (or, the mobility) can be expressed solely in terms of the first and second moments of $P(g)$, i.e.,
\begin{eqnarray}\label{ch4-D-inf-final_before}
D_{\infty}(\rho) &=& \frac{1}{4} - \frac{1}{4}\frac{\partial}{\partial \rho}\left[ \rho \langle g \rangle\right], \\
\chi_\infty(\rho) &=& \frac{\rho}{4} \left[ a(\rho) + \langle g^{2} \rangle\right].
\label{ch4-chi-inf-final_before}
\end{eqnarray}
Note that, by using the definition $\langle g \rangle = 1/\rho -1$, we can immediately obtain $D_{\infty}(\rho)$. On the other hand, the mobility $\chi_\infty(\rho)$ depends on the occupation probability $a(\rho)=1 - P(0)$ and second moment of the gap $\langle g^{2} \rangle$. Now, by using an independent-interval approximation (IIA) and then calculating the occupation probability and the second moment of gap distribution through the generating function for $P(g)$ \cite{Barma_PRL_1998}, we obtain
\begin{eqnarray}
    a[\tilde{\rho}(\rho)] &=& \frac{\tilde{\rho}(1 - \tilde{\rho})}{1 + \tilde{\rho}}, \\
    \langle g^{2} (\rho) \rangle &=& \frac{\tilde{\rho}[1 + a[\tilde{\rho}(\rho)]]}{1 - a[\tilde{\rho}(\rho)] - 2 \tilde{\rho}},
\end{eqnarray}
as a function of global density ${\rho}$, where we have $\tilde{\rho}=1/\rho -1$. Substituting the above two equations in Eq.~\eqref{ch4-chi-inf-final_before}, we finally obtain the mobility $\chi_{\infty}(\rho)$ as a function of global density. In the following equations, we write the explicit analytic expressions of the two density-dependent transport coefficients $-$ the bulk-diffusion coefficient and the mobility for lattice gases with infinite-ranged hopping, 
\begin{eqnarray}
\label{ch4-D-inf-final}
D_{\infty}(\rho) &=& \frac{1}{2}, \\
\chi_\infty(\rho) &=& \frac{\rho(1-\rho)(2\rho^2 - 2\rho + 1)}{2(2\rho^2-1)}.
\label{ch4-chi-inf-final}
\end{eqnarray}
Notably, unlike the previous case of finite-ranged hopping, the mobility $\chi_{\infty}(\rho)$ now has a singularity at $\rho=\rho_c=1/\sqrt{2}$. This kind of instability (divergence) in the collective particle mobility gives rise to a nonequilibrium phase transition in the system: for $\rho > \rho_c$, the system is in the normal or the disordered phase, while in the other limit of $\rho < \rho_c$, the system spontaneously phase separates into a critical fluid at density $\rho_c$ and a macroscopic vacancy or gap cluster having a size $(\rho_c-\rho)L$. It is therefore quite instructive to study the current fluctuation in more details at the critical density itself. Below, we first verify our analytical findings with numerical simulations at the normal phase, i.e., for $\rho > \rho_c$, and subsequently explore the most intriguing behavior of the current fluctuations at the critical point $\rho=\rho_c$.

\paragraph{Verification in the normal phase, $\rho > \rho_c$:} In the left panel of Fig.\ref{fig:ch4_bcf_linf}, we validate the time-integrated bond-current fluctuations $\langle Q^{2}(T) \rangle$ derived in Eqs.\eqref{ch4-bond-current-fluc} and \eqref{ch4-cf_limit} in the normal phase by plotting numerically obtained $\langle Q^{2}(T) \rangle$ against observation time $T$ across various densities, $\rho=0.75$, $0.8$, $0.85$, and $0.9$. Initially, the data points exhibit diffusive growth, transitioning to subdiffusive behavior ($\sqrt{T}$) before returning to diffusive growth, consistent with Eq.~\eqref{ch4-cf_limit}. To complement the simulation data, we also plot theoretical lines derived from Eq.\eqref{ch4-bond-current-fluc}, substituting transport coefficients from Eqs.\eqref{ch4-D-inf-final} and \eqref{ch4-chi-inf-final}, and solving for $\mathcal{F}_{n}$ using the numerically obtained steady-state gap distribution $P(g)$ in Eq.\eqref{ch4-f_{q}}. Notably, we observe a significant agreement between simulation data and theoretical curves, collectively validating Eqs.~\eqref{ch4-bond-current-fluc} and \eqref{ch4-cf_limit} within the normal phase of infinite-ranged hopping.

To verify the scaling form as given in Eq. \eqref{ch4-bond-current-fluc-3}, which should be valid in the disordered phase, we now proceed to plot the numerically derived scaled current fluctuation, expressed as $D_\infty \langle Q^{2}(T) \rangle/2\chi_\infty L$, against the scaled hydrodynamic time $y=D_\infty T/L^{2}$ in the right panel of Fig.\ref{fig:ch4_bcf_linf}. For this purpose, we consider a series of densities: $\rho=0.75$, $0.8$, $0.85$, and $0.9$. Remarkably, we observe that the simulation data points collapse onto one another exceptionally well, adhering closely to the theoretically derived scaling function $\mathcal{W}(y)$ (represented by the solid red line) as described in Eq.~\eqref{ch4-cf_scaling_function}. Furthermore, we note a transition in both the scaling function and the collapsed data points from $\sqrt{y}$ growth at smaller times to linear or $y$ behavior at significantly larger times. Consequently, the findings in simulations indeed confirm the validity of Eqs.~\eqref{ch4-bond-current-fluc-3}, \eqref{ch4-cf_scaling_function}, and \eqref{ch4-cf_scaling_function_limit} in the disordered phase.

We furthermore validate the derived fluctuation-response relation in Eq.~\eqref{ch4-I_fluctuation1} by numerically evaluating the space-time integrated current fluctuation $\langle Q^{2}_{tot}(L, T)\rangle$ in the steady state. For $\langle Q^{2}_{tot}(L, T)\rangle$, we perform simulations by using a system size of $L=1000$ and an observation time $T=1000$ and plot the scaled variance $\langle Q_{tot}^{2}(L, T) \rangle/2LT$ (represented by points) as a function of global density $\rho$ in Fig.\ref{fig:ch4_tcf_linf}. We also compare the numerical data points with the analytical solution of $2\chi_{\infty}(\rho)$ derived in Eq. \eqref{ch4-chi-inf-final}, shown as lines. Once again, we observe an excellent agreement between all simulation data points and theoretical lines, confirming the validity of Eq.~\eqref{ch4-I_fluctuation1} in the disordered phase.

\paragraph{Current fluctuation at the critical point} $\rho=\rho_c$:
As previously mentioned, while the bulk-diffusion coefficient $D_{\infty}(\rho_c)$ at the transition point remains constant and finite, the mobility, $\chi_{\infty}(\rho_c)$ however, diverges in the thermodynamic limit, leading to an anomalous scaling of the current fluctuations. At this stage, it would be quite interesting to inquire how $\chi_{\infty}(\rho_c, L)$ grows with system sizes $L$ when they are large but finite. Given that, at the critical point, the gap distribution function has the scaling form $P(g) \sim g^{-5/2} e^{-g/g_{0}}$ with $g_0 \propto L^{2/3}$ \cite{Rajesh_PRE_2001}, we can now incorporate the finite-size scaling into Eq.~\eqref{ch4-chi-inf-final_before} and we immediately obtain
\begin{eqnarray}
\label{chi_diverge_L}
\chi_{\infty}(\rho_c, L) \sim \langle g^{2} \rangle \sim L^{1/3}.
\end{eqnarray}
That is, at the critical point $\rho=\rho_c$, the mobility, or equivalently the scaled variance of space-time-integrated current across the entire system [see Eq.\eqref{ch4-I_fluctuation1}], diverges as $L^{1/3}$ in the large system size limit.
 Moreover, as the bulk-diffusion coefficient $D_{\infty}(\rho_c)$ still remains finite (constant), the scaling form as in Eq.~\eqref{ch4-bond-current-fluc-3} suggests that, at the critical point, 
  the variance $\langle Q^{2} (L, T) \rangle$ of time-integrated bond current should exhibit anomalous characteristics. then, using the finite-size scaling Eq. \eqref{chi_diverge_L}, we have $\langle Q^{2} (L, T) \rangle \sim L^{4/3}$ with $y_c=T/L^2$ fixed. 
In other words, we now have the following modified
  scaling form for the variance of time-integrated bond current at the critical point $\rho=\rho_c$:
\begin{eqnarray}
\label{scaling_ansatz_rhoc}
\langle Q^{2}(L, T) \rangle_c = \langle Q^{2}(L, T) \rangle \simeq L^{4/3} \mathcal{W}_{c}\left(\frac{T}{L^{2}} \right).
\end{eqnarray}
\begin{figure}[tpb]
           \centering
\includegraphics[width=0.75\linewidth]{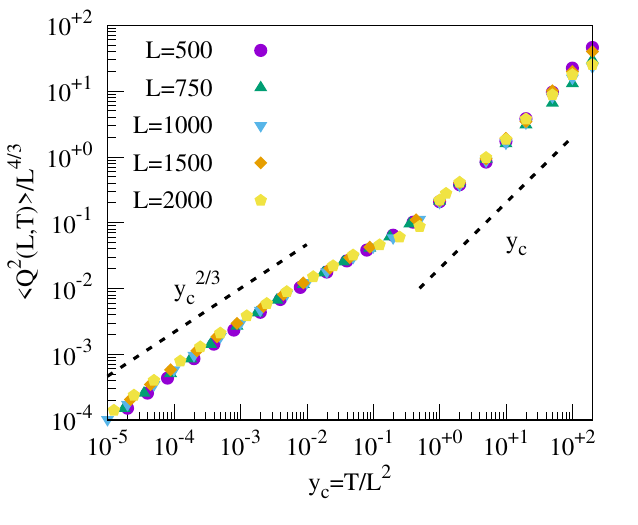}
         \caption{\small{We plot the scaled time-integrated bond-current fluctuation $\langle Q(L, T)^{2} \rangle/L^{4/3}$, as a function of the scaled time $y_c=T/L^{2}$ at the critical point (i.e., $l_0 \rightarrow \infty$ and $\rho=\rho_c$). The collapsed data points are shown to exhibit a initial-time subdiffusive growth $\sim y_{c}^{2/3}$, followed by a diffusive or linear growth $\sim y_c$ at large times.}}
          \label{fig:ch4_bcf_rho_c}
 \end{figure}
 We can now immediately calculate the asymptotic form of the scaling function by using the following arguments. We expect that, in the intermediate time regime ($ 1 \ll T \ll L^{2}$), the variance of time-integrated bond current should exhibit growth $\langle Q^{2}\rangle \sim T^{\nu}$, with an exponent $\nu$ and the prefactor being  $L$-independent; this temporal growth law should then cross over to a diffusive (linear) one, but now the prefactor being $L$-dependent, i.e., $\langle Q^{2}\rangle \sim T/L^{\delta}$, at large times, where $\delta$ is another exponent. Interestingly, the determination of the exponents $\nu$ and $\delta$ is now possible from the scaling ansatz Eq.~\eqref{scaling_ansatz_rhoc}. Using the fact that $\langle Q^{2}(L, T) \rangle$ is $L$-independent for $y_{c} \rightarrow 0$, we immediately obtain $\lim_{y_{c} \rightarrow 0} \mathcal{W}_{c}(y_c) \sim y_c^{2/3}$ and thus the exponent $\nu=2/3$. Moreover, the diffusive growth of $\langle Q^{2}(L, T) \rangle$ in the large $y_c$ limit suggests $\lim_{y_{c} \rightarrow \infty} \mathcal{W}_{c}(y_c) \sim y_c$ and the exponent $\delta=2/3$. Therefore, the asymptotic behavior of the scaling function at critical density $\rho=\rho_c$ is given by
\begin{eqnarray} 
 \mathcal{W}_{c}\left(y_{c}\right) \sim 
\left\{
\begin{array}{ll}
\vspace{0.15 cm}
  y_{c}^{2/3},            ~~~  & {\rm for}~ y_{c} \ll 1 , \\
\vspace{0.15 cm}
 y_c,            ~~~  & {\rm for}~ y_{c} \gg 1. \\
\end{array}
\right.
\label{ch4-cf_scaling_function_limit_criticality}
\end{eqnarray}
Now, using Eq.~\eqref{ch4-cf_scaling_function_limit_criticality}  in Eq.~\eqref{scaling_ansatz_rhoc}, we immediately obtain the limiting behaviors of the variance of time-integrated bond current $\langle Q^{2}(L, T) \rangle$ as a function of time $T$ at the critical density $\rho=\rho_c$ as follows:
\begin{eqnarray} 
 \langle Q^{2}(L, T) \rangle \sim 
\left\{
\begin{array}{ll}
\vspace{0.15 cm}
  T^{2/3},            ~~~  & {\rm for}~~~ 1 \ll D_{\infty}T \ll L^{2} , \\
\vspace{0.15 cm}
 T/L^{2/3},            ~~~  & {\rm for}~~~ D_{\infty}T \gg L^{2}. \\
\end{array}
\right.
\label{ch4-cf_limit_criticality}
\end{eqnarray}
By comparing the above limiting behavior with that in the normal phase (as depicted in Eq.~\eqref{ch4-cf_limit}), we immediately observe that there is an enhancement in the current fluctuation at criticality: in the moderate time regime, the variance of time-integrated bond current increases as $T^{2/3}$ instead of the $\sqrt{T}$ growth observed in the normal phase. Subsequently, there is a crossover to diffusive growth, albeit with a system-size dependent prefactor $1/L^{2/3}$ instead of $1/L$ as seen in the disordered phase.  
Notably, the above arguments are based on the scaling ansatz in Eq.~\eqref{scaling_ansatz_rhoc}, which we directly verify next through simulations. In  Fig. \ref{fig:ch4_bcf_rho_c}, we plot the numerically obtained scaled variance of the time-integrated bond current $\langle Q^{2}(L, T) \rangle/L^{4/3}$ as a function of the scaled (hydrodynamic) time $y_c=T/L^{2}$ at the critical density $\rho=\rho_c=1/\sqrt{2}$ for various system sizes $L= 500$, $750$, $1000$, $1500$, and $2000$. We observe that the data points collapse onto each other quite well, and the growth laws in the small and large time limits are indeed consistent with Eq.~\eqref{ch4-cf_scaling_function_limit_criticality}, as compared with the red-dotted lines. The above observations thus verify the scaling ansatz Eq.~\eqref{scaling_ansatz_rhoc}.

\section{Summary and concluding remarks}
\label{Sec:ch5-conclusion}

In this paper, we study time-dependent properties of hardcore lattice gases with extended-ranged hopping in terms of current fluctuations, where particles perform symmetric hopping within a finite range $l_0$ on a ring of $L$ sites, with total particle number being conserved. We specifically consider two cases with the hopping range $l_0 = 2$ and $l_0 \to \infty$, both of which, albeit having a nontrivial spatial structure, are analytically tractable. In the latter case, the system, which is related to a previously studied conserved-mass aggregation model \cite{Barma_PRL_1998}, is known to undergo a nonequilibrium clustering or condensation transition, which occurs below a critical density $0 < \rho_c < 1$. 
In the present work, we \textcolor{black}{theoretically characterize the large-scale dynamic fluctuation properties of various time-dependent quantities, such as time- and space-integrated currents in the systems, with both finite- and infinite-ranged hopping; in the latter case, we also dynamically characterize the phase transition observed in the system. }
 Using a closure scheme, we analytically study the variance of time-integrated bond current (as well as space-time-integrated current across the entire system) and the density-dependent Onsager coefficient, or the collective particle mobility, as defined in Eq. \eqref{var1}.
 We explicitly calculate the closed-form analytic expressions of these quantities and verify the theoretical results through extensive Monte Carlo simulations, which are in excellent agreement with our theory.

 The large-scale properties of diffusive systems are in general described by the  two macroscopic transport coefficients - the bulk-diffusion coefficient \( D(\rho) \) and the mobility \( \chi(\rho) \)  \cite{Bertini_PRL2001, Derrida_PRL2004}. 
Indeed, on a macroscopic level, the onset of clustering manifests itself through an instability, or a singularity, in the transport properties of the system, i.e., the transport coefficients can be nonanalytic at the critical point. In equilibrium systems, such as the paradigmatic Ising lattice gases with particle number conserved,  the system, on approaching the critical point, dynamically \textit{slows down,} and the relaxation rate, or equivalently, the bulk-diffusion coefficient $D(\rho)$, vanishes in the thermodynamic limit \cite{Kawasaki_csd, Binder_CSD}. 
Recently, this particular phenomenological theory has been extended to nonequilibrium systems, e.g., clustering phenomena in interacting self-propelled particles \cite{Marchetti_PRL_2012, Baskaran_PRL_2013, Tailleur_PRL_2018} undergoing a motility-induced phase separation (MIPS) \cite{RTP_tailleur, MIPS_2015} and diffusivity-edge induced condensation transition in scalar active matter \cite{Golestanian_PRE_2019, Berx_EPL_2023}. Thus, at the critical point of a clustering transition, one would expect, by taking a cue from equilibrium, a critical slowing down, which has indeed been considered to be a universal mechanism, even for clustering in a nonequilibrium setting. However, in this paper, we argue that this is not always the case and substantiate our claim by providing a counterexample of lattice gases having infinite-ranged hopping. 
Indeed, quite contrary to the expectations, we show that, while the density relaxation in such systems remains diffusive far and near criticality (thus the bulk-diffusion coefficient being strictly nonzero), remarkably the Onsager transport coefficient or the mobility, $\chi(\rho) \sim (\rho - \rho_c)^{-1}$ {\it diverges} at the critical point in the thermodynamic limit $-$ a new paradigm, which we call mobility-driven clustering.

For simplicity, we consider a class of models where the hop length is a random variable ${\rm min}(g,l)$, where the length $l$ is drawn from a distribution $\phi(l) = \alpha \delta_{l,1} + \beta \delta_{l,l_0}$, $g$ is the inter-particle gap in the hopping direction and $l_0$ is the characteristic hopping range. Interestingly, these models are amenable to analytical calculations for $l_0=2$ and $l_0 \to \infty$.
For $l_0=2$, we explicitly calculate the variance of time-integrated bond current as a function of time $t$, density $\rho$, and system size $L$: The variance exhibits characteristics typical of diffusive systems with a single conservation law \cite{Derrida-Sadhu-JSTAT2016, Hazra_Jstat_2024}. In this case, there is no phase transition for $0 < \rho < 1$, and the temporal growth of the variance of time-integrated bond current exhibits three distinct regimes as follows. Initially, in regime I ($t \ll 1/D$), the fluctuation grows proportional to time $t$. Then, it crosses over to regime II ($1 \ll t \ll L^2$), where the variance exhibits subdiffusive $\sqrt{t}$ growth. Finally, it crosses over to a linear temporal-growth regime III ($t \gg L^2$), where $\langle Q^{2}(t) \rangle \sim t/L$. Notably, the prefactors of $\langle Q^{2}(t) \rangle$ in regimes II and III can be expressed as the density-dependent macroscopic transport coefficients, $-$ the bulk-diffusion and the mobility. Interestingly, upon suitable finite-size scaling, $\langle Q^{2}(t) \rangle$ is shown to satisfy a scaling law, which has been theoretically characterized.

In the most interesting scenario of infinite-range hopping (IRH), below a critical density $\rho_c$, the system undergoes a phase transition from a disordered fluid phase to a translation-symmetry-broken ordered phase with a macroscopic condensate of vacancies or holes \cite{Barma_PRL_1998}.
Indeed, understanding clustering phenomena dynamically in such systems requires probing transport properties of the systems, involving time-dependent quantities like cumulative currents across a bond and the entire system in finite, large time intervals.
In the present study, we demonstrate that the clustering or condensation transition indeed manifests itself through anomalous current fluctuations at the critical point $\rho=\rho_c$. On the moderately large timescale ($1 \ll t \ll L^2$, regime II), the variance of time-integrated bond current is found to exhibit a subdiffusive growth $\langle Q^{2}(t) \rangle \sim t^{\nu}$ with an exponent $ \nu = 2/3$. This is strikingly different from the typical subdiffusive growth of the variance of bond current in the disordered phase discussed above, where the growth is characterized by an exponent of $\nu=1/2$. 
Finally, on a very large timescale ($t \gg L^2$, regime III), the time-integrated bond-current fluctuation displays a linear growth. While this particular growth law resembles that found in the disordered phase, the unique nature of the current fluctuations is related to the anomalous system-size dependence of the prefactor of the linear growth law: at the critical point, $\langle Q^{2}(t) \rangle$ varies as $t/L^{2/3}$, which differs from the typical $t/L$ behavior in the disordered phase.

Also, it is worth comparing our results with those obtained in Ref.~\cite{Chakraborty_PRE_2024_ZRP} for an equilibrium system $-$ the symmetric zero-range processes (ZRPs), which, under certain conditions on the (mass-dependent) hopping rates, undergo a condensation transition and exhibit, like in the Ising lattice gases \cite{Bhattacharjee_1985}, critical slowing down. In the above-mentioned regime III, the temporal growth of the variance $\langle Q^{2}(t) \rangle_c$ of time-integrated bond current at the critical point is linear ($\sim t$) in both the ZRPs and the lattice gases with infinite-ranged hopping (IRH). The linear growth is quite similar to that observed in a typical diffusive system with a single conserved quantity such as density \cite{Derrida-Sadhu-JSTAT2016, Hazra_Jstat_2024}. But, as mentioned before,  in lattice gases with IRH, the system-size-dependent prefactor in the temporal growth of the variance at the critical point are  anomalous (diverging) and differ from that in ZRPs, where the  prefactor is finite. 
Moreover, while, in the (equilibrium) ZRPs, the crossover time between regimes II and III at the critical point increases anomalously with system size as $L^z$, where the dynamic exponent $z > 2$ (thus implying critical slowing down), but, in the (nonequilibrium) lattice gases with IRH, it varies as $L^2$ (diffusive). 
 On the other hand, at the intermediate time scales (regime II), there is a striking resemblance, though, between in- and out-of-equilibrium current fluctuations at the critical point: the temporal growths of the variance of time-integrated bond current in both cases are larger ($\nu > 1/2$) than that ($\nu = 1/2$) observed in the disordered phases.

Overall, our theoretical findings clearly indicate a greatly enhanced growth of time-integrated bond-current fluctuations upon approaching the critical point for a nonequilibrium condensation transition in lattice gases with infinite-ranged hopping. Interestingly, the enhanced fluctuations across the individual bonds collectively cause the scaled space-time-integrated current fluctuation across the entire system to also exhibit an anomalous behavior: near criticality, the latter one actually \textit{diverges} in the thermodynamic limit as $(\rho - \rho_c)^{-1}$; this is in contrast to the behavior in the disordered phase, where the scaled space-time-integrated current fluctuation is in fact finite. 
In the equilibrium condensation transition, the bulk-diffusion coefficient vanishes at the critical point, thus leading to an anomalously slow relaxation in the systems; in contrast, the dynamics in nonequilibrium setting as considered here remain diffusive throughout, i.e., {\it there is no critical slowing down}, and the phase transition is induced by the diverging current fluctuations. \textcolor{black}{Note that, in the case of infinite-ranged hopping, the typical hop length is cut-off by the typical gap size $\sqrt{\langle g^2 \rangle}$, which, though finite at large densities, diverges near the critical point $\rho=\rho_c$,
thus allowing particles to traverse large distance and to generate large instantaneous currents. This physically explains why there is anomalously large current fluctuations in the system at the critical point.}
Indeed, in the context of clustering phenomena, which are ubiquitous in nature, this mechanism leading to the clustering in infinite-ranged lattice gases is quite unique. We believe our findings could offer a new perspective on the theoretical understanding of clustering transitions in various other out-of-equilibrium systems.

\section*{Acknowledgement}

\textcolor{black}{We are grateful to Deepak Dhar and Kavita Jain for useful discussions, critical reading of the manuscript, and insightful comments. We also gratefully acknowledge Sanjay Puri for useful discussions. }

\section*{References}
\bibliographystyle{unsrt}
\bibliography{main_jstat.bib}

\begin{thebibliography}{10}

\bibitem{Spohn_2012}
Herbert Spohn.
\newblock {\em Large Scale Dynamics of Interacting Particles}.
\newblock Springer Berlin, Heidelberg, 1991.

\bibitem{Marro_Dickman_1999}
Joaquin Marro and Ronald Dickman.
\newblock {\em Nonequilibrium Phase Transitions in Lattice Models}.
\newblock Collection Alea-Saclay: Monographs and Texts in Statistical Physics.
  Cambridge University Press, 1999.

\bibitem{Halperin_RMP}
P.~C. Hohenberg and B.~I. Halperin.
\newblock Theory of dynamic critical phenomena.
\newblock {\em Rev. Mod. Phys.}, 49:435--479, Jul 1977.

\bibitem{Kawasaki_csd}
Kyozi Kawasaki.
\newblock Diffusion constants near the critical point for time-dependent ising
  models. i.
\newblock {\em Phys. Rev.}, 145:224--230, May 1966.

\bibitem{Binder_CSD}
K.~Binder, D.~Stauffer, and H.~M\"uller-Krumbhaar.
\newblock Theory for the dynamics of clusters near the critical point. i.
  relaxation of the glauber kinetic ising model.
\newblock {\em Phys. Rev. B}, 12:5261--5287, Dec 1975.

\bibitem{Bhattacharjee_1985}
Jayanta~K. Bhattacharjee.
\newblock Dynamic critical exponent for the ferromagnetic transition.
\newblock {\em Hyperfine Interactions}, 25(1):427--434, Nov 1985.

\bibitem{MFT-RMP2015}
Lorenzo Bertini, Alberto De~Sole, Davide Gabrielli, Giovanni Jona-Lasinio, and
  Claudio Landim.
\newblock Macroscopic fluctuation theory.
\newblock {\em Rev. Mod. Phys.}, 87:593--636, Jun 2015.

\bibitem{Derrida_PRL2004}
T.~Bodineau and B.~Derrida.
\newblock Current fluctuations in nonequilibrium diffusive systems: An
  additivity principle.
\newblock {\em Phys. Rev. Lett.}, 92:180601, May 2004.

\bibitem{tanmoy_2020}
Tanmoy Chakraborty, Subhadip Chakraborti, Arghya Das, and Punyabrata Pradhan.
\newblock Hydrodynamics, superfluidity, and giant number fluctuations in a
  model of self-propelled particles.
\newblock {\em Phys. Rev. E}, 101:052611, May 2020.

\bibitem{Derrida:1998}
Bernard Derrida and Joel~L. Lebowitz.
\newblock Exact large deviation function in the asymmetric exclusion process.
\newblock {\em Phys. Rev. Lett.}, 80:209--213, Jan 1998.

\bibitem{Derrida:2007}
Bernard Derrida.
\newblock Non-equilibrium steady states: fluctuations and large deviations of
  the density and of the current.
\newblock {\em Journal of Statistical Mechanics: Theory and Experiment},
  2007(07):P07023, jul 2007.

\bibitem{Derrida-PRE2008}
C.~Appert-Rolland, B.~Derrida, V.~Lecomte, and F.~van Wijland.
\newblock Universal cumulants of the current in diffusive systems on a ring.
\newblock {\em Phys. Rev. E}, 78:021122, Aug 2008.

\bibitem{Derrida_2009_JSTAT_1}
Bernard Derrida and Antoine Gerschenfeld.
\newblock Current fluctuations of the one-dimensional symmetric simple
  exclusion process with step initial condition.
\newblock {\em Journal of Statistical Physics}, 136(1):1--15, 2009.

\bibitem{Imparato:2009}
A.~Imparato, V.~Lecomte, and F.~van Wijland.
\newblock Equilibriumlike fluctuations in some boundary-driven open diffusive
  systems.
\newblock {\em Phys. Rev. E}, 80:011131, Jul 2009.

\bibitem{Krapivsky:2012}
P.~L. Krapivsky and Baruch Meerson.
\newblock Fluctuations of current in nonstationary diffusive lattice gases.
\newblock {\em Phys. Rev. E}, 86:031106, Sep 2012.

\bibitem{Gorissen:2009}
Mieke Gorissen, Jef Hooyberghs, and Carlo Vanderzande.
\newblock Density-matrix renormalization-group study of current and activity
  fluctuations near nonequilibrium phase transitions.
\newblock {\em Phys. Rev. E}, 79:020101, Feb 2009.

\bibitem{Harris:2005}
R~J Harris, A~Rákos, and G~M Schütz.
\newblock Current fluctuations in the zero-range process with open boundaries.
\newblock {\em Journal of Statistical Mechanics: Theory and Experiment},
  2005(08):P08003, aug 2005.

\bibitem{Kipnis:1982}
C.~Kipnis, C.~Marchioro, and E.~Presutti.
\newblock Heat flow in an exactly solvable model.
\newblock {\em Journal of Statistical Physics}, 27(1):65--74, Jan 1982.

\bibitem{Basile:2006}
Giada Basile, C\'edric Bernardin, and Stefano Olla.
\newblock Momentum conserving model with anomalous thermal conductivity in low
  dimensional systems.
\newblock {\em Phys. Rev. Lett.}, 96:204303, May 2006.

\bibitem{Hurtado:2009}
Pablo~I Hurtado and Pedro~L Garrido.
\newblock Current fluctuations and statistics during a large deviation event in
  an exactly solvable transport model.
\newblock {\em Journal of Statistical Mechanics: Theory and Experiment},
  2009(02):P02032, feb 2009.

\bibitem{GrandPre:2018}
Trevor GrandPre and David~T Limmer.
\newblock Current fluctuations of interacting active brownian particles.
\newblock {\em Physical Review E}, 98(6):060601, 2018.

\bibitem{Banerjee:2020}
Tirthankar Banerjee, Satya~N. Majumdar, Alberto Rosso, and Gr\'egory Schehr.
\newblock Current fluctuations in noninteracting run-and-tumble particles in
  one dimension.
\newblock {\em Phys. Rev. E}, 101:052101, May 2020.

\bibitem{Chakraborty_PRE_2024_RTP_II}
Tanmoy Chakraborty and Punyabrata Pradhan.
\newblock Time-dependent properties of run-and-tumble particles. ii. current
  fluctuations.
\newblock {\em Phys. Rev. E}, 109:044135, Apr 2024.

\bibitem{Jose_PRE_2023}
Stephy Jose, Alberto Rosso, and Kabir Ramola.
\newblock Generalized disorder averages and current fluctuations in run and
  tumble particles.
\newblock {\em Phys. Rev. E}, 108:L052601, Nov 2023.

\bibitem{Bertini_PRL2001}
L.~Bertini, A.~De~Sole, D.~Gabrielli, G.~Jona-Lasinio, and C.~Landim.
\newblock Fluctuations in stationary nonequilibrium states of irreversible
  processes.
\newblock {\em Phys. Rev. Lett.}, 87:040601, Jul 2001.

\bibitem{Bertini:2002}
L.~Bertini, A.~De~Sole, D.~Gabrielli, G.~Jona-Lasinio, and C.~Landim.
\newblock Macroscopic fluctuation theory for stationary non-equilibrium states.
\newblock {\em Journal of Statistical Physics}, 107(3):635--675, May 2002.

\bibitem{Evans_PRL_1998}
M.~R. Evans, Y.~Kafri, H.~M. Koduvely, and D.~Mukamel.
\newblock Phase separation in one-dimensional driven diffusive systems.
\newblock {\em Phys. Rev. Lett.}, 80:425--429, Jan 1998.

\bibitem{Gerschenfeld_EPL_2011}
A.~Gerschenfeld and B.~Derrida.
\newblock Current fluctuations at a phase transition.
\newblock {\em Europhysics Letters}, 96(2):20001, sep 2011.

\bibitem{Remlein_JCP_2024}
Benedikt Remlein and Udo Seifert.
\newblock Nonequilibrium fluctuations of chemical reaction networks at
  criticality: The schlögl model as paradigmatic case.
\newblock {\em The Journal of Chemical Physics}, 160(13):134103, 04 2024.

\bibitem{Fiore:2021}
C.~E. Fiore, Pedro~E. Harunari, C.~E.~Fern\'andez Noa, and Gabriel~T. Landi.
\newblock Current fluctuations in nonequilibrium discontinuous phase
  transitions.
\newblock {\em Phys. Rev. E}, 104:064123, Dec 2021.

\bibitem{Nguyen_JCP_2018}
Basile Nguyen, Udo Seifert, and Andre~C. Barato.
\newblock Phase transition in thermodynamically consistent biochemical
  oscillators.
\newblock {\em The Journal of Chemical Physics}, 149(4):045101, 07 2018.

\bibitem{ptaszynski_PRE_2024}
Krzysztof Ptaszy\ifmmode~\acute{n}\else \'{n}\fi{}ski and Massimiliano
  Esposito.
\newblock Critical heat current fluctuations in curie-weiss model in and out of
  equilibrium.
\newblock {\em Phys. Rev. E}, 111:034125, Mar 2025.

\bibitem{Anirban-PRE2023}
Anirban Mukherjee and Punyabrata Pradhan.
\newblock Dynamic correlations in the conserved manna sandpile.
\newblock {\em Phys. Rev. E}, 107:024109, Feb 2023.

\bibitem{Mukherjee_PRE_2024}
Anirban Mukherjee, Dhiraj Tapader, Animesh Hazra, and Punyabrata Pradhan.
\newblock Anomalous relaxation and hyperuniform fluctuations in center-of-mass
  conserving systems with broken time-reversal symmetry.
\newblock {\em Phys. Rev. E}, 110:024119, Aug 2024.

\bibitem{Chakraborty_PRE_2024_ZRP}
Tanmoy Chakraborty, Punyabrata Pradhan, and Kavita Jain.
\newblock Current fluctuations in the symmetric zero-range process below and at
  critical density.
\newblock {\em Phys. Rev. E}, 110:L052103, Nov 2024.

\bibitem{agranov_2023_scipost}
Tal Agranov, Sunghan Ro, Yariv Kafri, and Vivien Lecomte.
\newblock Macroscopic fluctuation theory and current fluctuations in active
  lattice gases.
\newblock {\em SciPost Physics}, 14(3):045, 2023.

\bibitem{Jose:2023}
Stephy Jose, Rahul Dandekar, and Kabir Ramola.
\newblock Current fluctuations in an interacting active lattice gas.
\newblock {\em Journal of Statistical Mechanics: Theory and Experiment},
  2023(8):083208, aug 2023.

\bibitem{Derrida-Sadhu-JSTAT2016}
Tridib Sadhu and Bernard Derrida.
\newblock Correlations of the density and of the current in non-equilibrium
  diffusive systems.
\newblock {\em Journal of Statistical Mechanics: Theory and Experiment},
  2016(11):113202, 2016.

\bibitem{Hazra_Jstat_2024}
Animesh Hazra, Anirban Mukherjee, and Punyabrata Pradhan.
\newblock Dynamic fluctuations of current and mass in nonequilibrium mass
  transport processes.
\newblock {\em Journal of Statistical Mechanics: Theory and Experiment},
  2024(8):083205, aug 2024.

\bibitem{Barma_PRL_1998}
Satya~N. Majumdar, Supriya Krishnamurthy, and Mustansir Barma.
\newblock Nonequilibrium phase transitions in models of aggregation,
  adsorption, and dissociation.
\newblock {\em Phys. Rev. Lett.}, 81:3691--3694, Oct 1998.

\bibitem{Landim_1998}
Claude Kipnis and Claudio Landim.
\newblock {\em Scaling limits of interacting particle systems}, volume 320.
\newblock Springer Science \& Business Media, 1998.

\bibitem{Mallick-PRE2014}
Chikashi Arita, P.~L. Krapivsky, and Kirone Mallick.
\newblock Generalized exclusion processes: Transport coefficients.
\newblock {\em Phys. Rev. E}, 90:052108, Nov 2014.

\bibitem{subhadip_PRE_2021}
Subhadip Chakraborti, Tanmoy Chakraborty, Arghya Das, Rahul Dandekar, and
  Punyabrata Pradhan.
\newblock Transport and fluctuations in mass aggregation processes:
  Mobility-driven clustering.
\newblock {\em Phys. Rev. E}, 103:042133, Apr 2021.

\bibitem{beck2010art}
Matthias Beck and Ross Geoghegan.
\newblock {\em The art of proof: basic training for deeper mathematics}.
\newblock Springer New York, NY, 2010.

\bibitem{Rajesh_PRE_2001}
R.~Rajesh and Satya~N. Majumdar.
\newblock Exact phase diagram of a model with aggregation and chipping.
\newblock {\em Phys. Rev. E}, 63:036114, Feb 2001.

\bibitem{Marchetti_PRL_2012}
Yaouen Fily and M.~Cristina Marchetti.
\newblock Athermal phase separation of self-propelled particles with no
  alignment.
\newblock {\em Phys. Rev. Lett.}, 108:235702, Jun 2012.

\bibitem{Baskaran_PRL_2013}
Gabriel~S. Redner, Michael~F. Hagan, and Aparna Baskaran.
\newblock Structure and dynamics of a phase-separating active colloidal fluid.
\newblock {\em Phys. Rev. Lett.}, 110:055701, Jan 2013.

\bibitem{Tailleur_PRL_2018}
Mourtaza Kourbane-Houssene, Cl\'ement Erignoux, Thierry Bodineau, and Julien
  Tailleur.
\newblock Exact hydrodynamic description of active lattice gases.
\newblock {\em Phys. Rev. Lett.}, 120:268003, Jun 2018.

\bibitem{RTP_tailleur}
J.~Tailleur and M.~E. Cates.
\newblock Statistical mechanics of interacting run-and-tumble bacteria.
\newblock {\em Phys. Rev. Lett.}, 100:218103, May 2008.

\bibitem{MIPS_2015}
Michael~E. Cates and Julien Tailleur.
\newblock Motility-induced phase separation.
\newblock {\em Annual Review of Condensed Matter Physics}, 6(1):219--244, 2015.

\bibitem{Golestanian_PRE_2019}
Ramin Golestanian.
\newblock Bose-einstein-like condensation in scalar active matter with
  diffusivity edge.
\newblock {\em Phys. Rev. E}, 100:010601, Jul 2019.

\bibitem{Berx_EPL_2023}
Jonas Berx, Aritra Bose, Ramin Golestanian, and Benoît Mahault.
\newblock Reentrant condensation transition in a model of driven scalar active
  matter with diffusivity edge.
\newblock {\em Europhysics Letters}, 142(6):67004, jun 2023.

\end{thebibliography}

\end{document}